\documentclass[preprint,showpacs,superscriptaddress,groupedaddress]{aastex}  

\usepackage{graphicx}
\usepackage{dcolumn}
\usepackage{bm}
\usepackage{amssymb}
\usepackage{graphicx}
\usepackage{dcolumn}
\usepackage{bm}
\usepackage{amsmath, amssymb, amsfonts, amsthm}

\newcommand\simlt{\lower.5ex\hbox{$\; \buildrel < \over \sim \;$}}
\newcommand\simgt{\lower.5ex\hbox{$\; \buildrel > \over \sim \;$}}

\begin{document}
\tighten
\title{Directed searches for broadband extended gravitational-wave emission in nearby energetic core-collapse supernovae}
\author{Maurice, H.P.M. Van Putten$^{a}$\footnote{corresponding author: mvp@sejong.ac.kr}  }
\affil{$^a$ Room 614, Astronomy and Space Science, Sejong University, 98 Gunja-Dong Gwangin-gu, Seoul 143-747, Korea}





\begin{abstract}
Core-collapse supernovae are factories of neutron stars and stellar mass black holes. Type Ib/c supernovae
stand out as potentially originating in relatively compact stellar binaries and their branching ratio of about 1\%
into long gamma-ray bursts. The most energetic events probably derive from central engines harboring
rapidly rotating black holes, wherein accretion of fall-back matter down to the Inner Most Stable Circular Orbit (ISCO) 
offers a window to {\em broadband extended gravitational-wave emission} (BEGE). To search for BEGE, we introduce a butterfly filter in time-frequency space by Time Sliced Matched Filtering. To analyze long epochs of data, we propose using coarse grained searches followed by high resolution searches on events of interest. We illustrate our proposed coarse grained search on two weeks of LIGO S6 data prior to SN 2010br $(z=0.002339)$ using a bank of up to 64 thousand templates of one second duration covering a broad range in chirp frequencies and bandwidth. Correlating events with signal-to-noise ratios $>6$ from the LIGO L1 and H1 detectors each reduces to a few events of interest. Lacking any further properties reflecting a common excitation by broadband gravitational radiation, we disregarded these as spurious. This new pipeline may be used to systematically search for long duration chirps in nearby core-collapse supernovae from robotic optical transient surveys using embarrassingly parallel computing. 
\end{abstract} 

\maketitle


\section{Introduction}

Cosmological gamma-ray bursts (GRBs) and core-collapse supernovae (CC-SNe) 
are the most extreme transients in the sky. The latter are quite frequent \citep[e.g.][]{cap15},
about once per fifty years in the Milky Way \citep{die06} and over once per decade in
nearby galaxies such as M51 ($D\simeq 8$ Mpc) and M82 $(D\simeq 4$ Mpc). 
CC-SNe of Type Ib/c have a branching ratio of about 1\% into normal long GRBs (LGRB) \citep{del03,van04,del06,gue07}. CC-SNe are generally factories of neutron stars and black holes. A relativistic inner engine of such kind gives a unique outlook on potentially powerful emissions in gravitational waves, that may be probed by upcoming gravitational wave observations by LIGO-Virgo \citep{abr92,ace06,ace07} and KAGRA \citep{som12,kag14}. 

Modern robotic optical surveys of the Local Transient Universe (LTU) such as P60 \citep{dro11} and
LOSS \citep{li11} provide an increasingly large number of nearby core-collapse supernovae with a yield of tens 
of events per year within a distance of about one hundred Mpc. Higher yields are expected from
Pan-STARRS \citep{pan11} and the Zwicky Transient Factory \citep{kul14,bel15}. At present,
SN2010br ($z=0.0023$) \citep{nev10,cho10} exemplifies a most nearby SNIb/c 
discovered by traditional means in the constellation Ursa Major.

Type Ib/c supernovae stand out as energetic events, that are
aspherical and radio loud \citep{maz05,tau09,mod14} featuring mildly relativistic ejecta \citep[e.g.][]{sod08,sod10,del10}. 
Formed in core-collapse of relatively massive progenitor stars stripped of its hydrogen and helium envelope, they may originate in compact stellar binaries with intra-day periods (e.g. \cite{pac98,del10,heo15} and references therein) or they have Wolf-Rayet progenitors \citep{woo93,woo06,sma09}. These considerations suggest that SNIb/c are engine driven.
If so by newly formed black holes, those with progenitors in compact binaries should be angular momentum rich
by conservation of angular momentum in collapse \citep{van04} that, in fact, may be near-extremal in producing LGRBs\citep{van15b}. Explosions driven by angular momentum-rich central engines provide a natural candidate to account for the observed powerful aspherical explosions \citep{bis70}, especially so with outflows taken to their relativistic limits \citep{mac99}. 

The SNII event SN1987A provided first-principle evidence for the formation of high density matter from its
$>$10MeV neutrino burst. From its current aspherical remnant, it must have been angular momentum
rich. This leaves us but one step away from emission in gravitational waves by non-axisymmetric mass motion,
from fall-back matter of the progenitor envelope. This outlook is especially relevant in the presence of
feedback by central engines onto matter falling in, notably so from rotating black holes \citep{van99,van03}. 
Powerful feedback derived from an angular momentum-rich energy reservoir may drive secular instabilities and sustain long duration emission \citep{van12}, much beyond what would be expected without in canonical core-collapse scenarios. 

Non-axisymmetric fall-back matter can hereby produce {\em ascending chirps} up to several hundred Hz \citep{pir07,lev15}, while non-axisymmetric ISCO waves can produce {\em descending chirps} down to several hundred Hz in process of feedback by the black hole \citep{van08}. As the latter takes place on the scale of the Schwarzschild radius of the system, the total energy radiated in gravitational waves can reach a fraction of order unity of the total spin energy of the central engine, i.e., a few tenths of a solar mass \citep{van01a}. This combined outlook points to broadband extended gravitational-wave emission (BEGE) from events producing black holes (Fig. 1),
that may further include gravitational wave emission from turbulent mass motion. Probing these events for the nature of their inner engine, therefore, requires a broadband detection algorithm sensitivity to both ascending and descending chirps. 

\begin{figure*}[h]
\centerline{\includegraphics[scale=0.7]{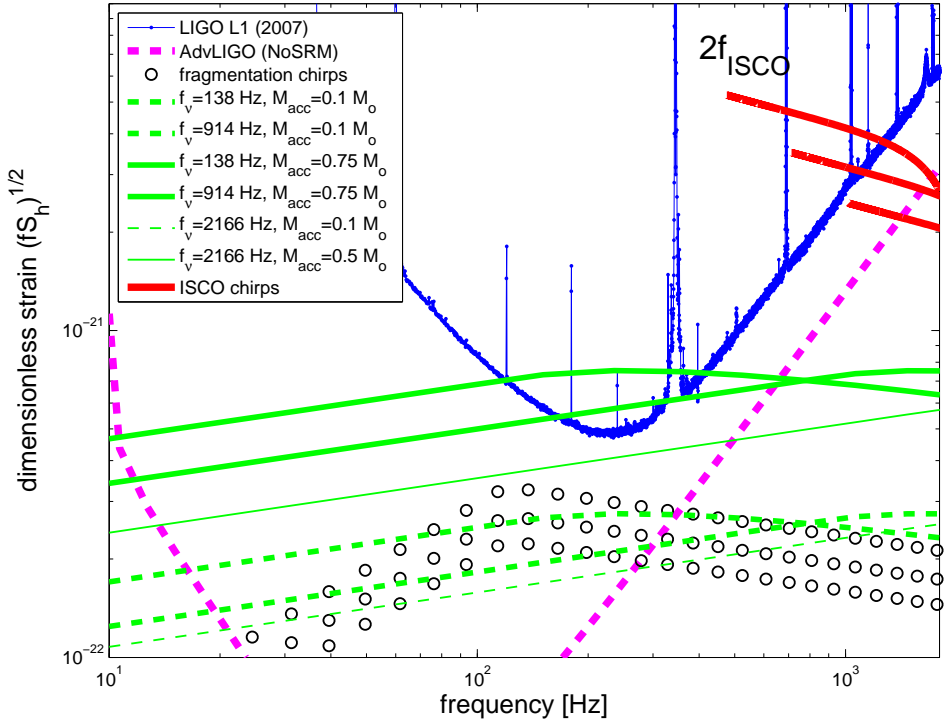}}
\caption{Overview of the characteristic gravitational wave strain $h_{char}(f)$ of quadrupole gravitational radiation at a frequency $f$ from accretion flows onto rotating black holes formed in core-collapse of massive stars at $D=100$ Mpc. The vertical distance to the dimensionless strain noise $h_n=\sqrt{fS_h}$ in LIGO S5 represents the theoretical limit of signal-to-noise ratio in matched filtering searches. Orientation-averaged signal strength of model curves shown are (a) broadband emission from non-axisymmetric accretion flows \citep[][{green solid and dashed curves}]{lev15}, (b) fragmentation chirps \citep[][{circles}]{pir07} and (c) chirps \citep[][{red solid curves}]{van08} at twice the orbital frequency $f_{ISCO}$ from non-axisymmetric ISCO waves stimulated by feedback from a rapidly rotating black hole. The red curves shown refer to a black hole mass $M=10M_\odot$ and (a) and $M=7,10$ and $15M_\odot$ in (b) and (c). (Adapted from \cite{van04,lev15}.) }
\label{fig:hchar}
\end{figure*}

For gravitational wave strain data from the ISCO around rapidly rotating black holes \citep{van08}, signal injection experiments demonstrate a sensitivity distance $D\simeq 100$ Mpc at advanced LIGO sensitivity for energetic Type Ib/c events with an output of a few tenths of a solar mass-energy in waves from the ISCO \citep{van01a,van11}. This sensitivity distance  gives an appreciable volume of the local universe with aforementioned event rate of Type Ib/c events. A similar sensitivity distance may hold for accretion flows with sufficiently rapid cooling \citep[][and references therein]{lev15}. In frequency, emission from the ISCO accurately carries detailed information on the size of the black hole as defined by the Kerr metric. Of particular interest, therefore, is identification of the associated evolution of the black hole spin in a secular change in gravitational wave frequency from matter at and about the ISCO \citep{van11a}.
Core-collapse supernovae during upcoming advanced LIGO-Virgo and KAGRA observations will provide us with the means for a test this hypothesis, provided they are captured sufficiently nearby and preferably directed a well-sampled optical supernova light curve.

Here, we consider SN 2010br as an illustrative example covered by the sixth science run LIGO \citep{val14} around the time of its discovery on April 10 2010.  Its light curve is sparsely sampled about the time of its discovery, however, and it shows an absolute magnitude of $M_R \simeq -12.3$ ($m_R\simeq 17.7$ using a distance modulus of 30). It is therefore intrinsically faint or discovery was late after maximal luminosity. Despite these caveats, SN 2010br provides a rare opportunity, given the overall event rate of Type Ib/c supernovae of about one hundred per year within a distance of 100 Mpc. SN 2010br therefore poses an interesting, rare but challenging example motivating the present development of a dedicated pipeline to search for BEGE from nearby core-collapse events, in preparation of future systematic approaches based on triggers provided by robotic optical transient surveys of the Local Universe \citep{heo15}. It exemplifies more broadly directed searches for unmodeled burst and transient sources \citep{aas15}. For concreteness, we selected a period of two to four weeks before its discovery, that conceivably covers the true time-of-onset $t_0$. 

\begin{figure}[h]
\centerline{\includegraphics[scale=0.46]{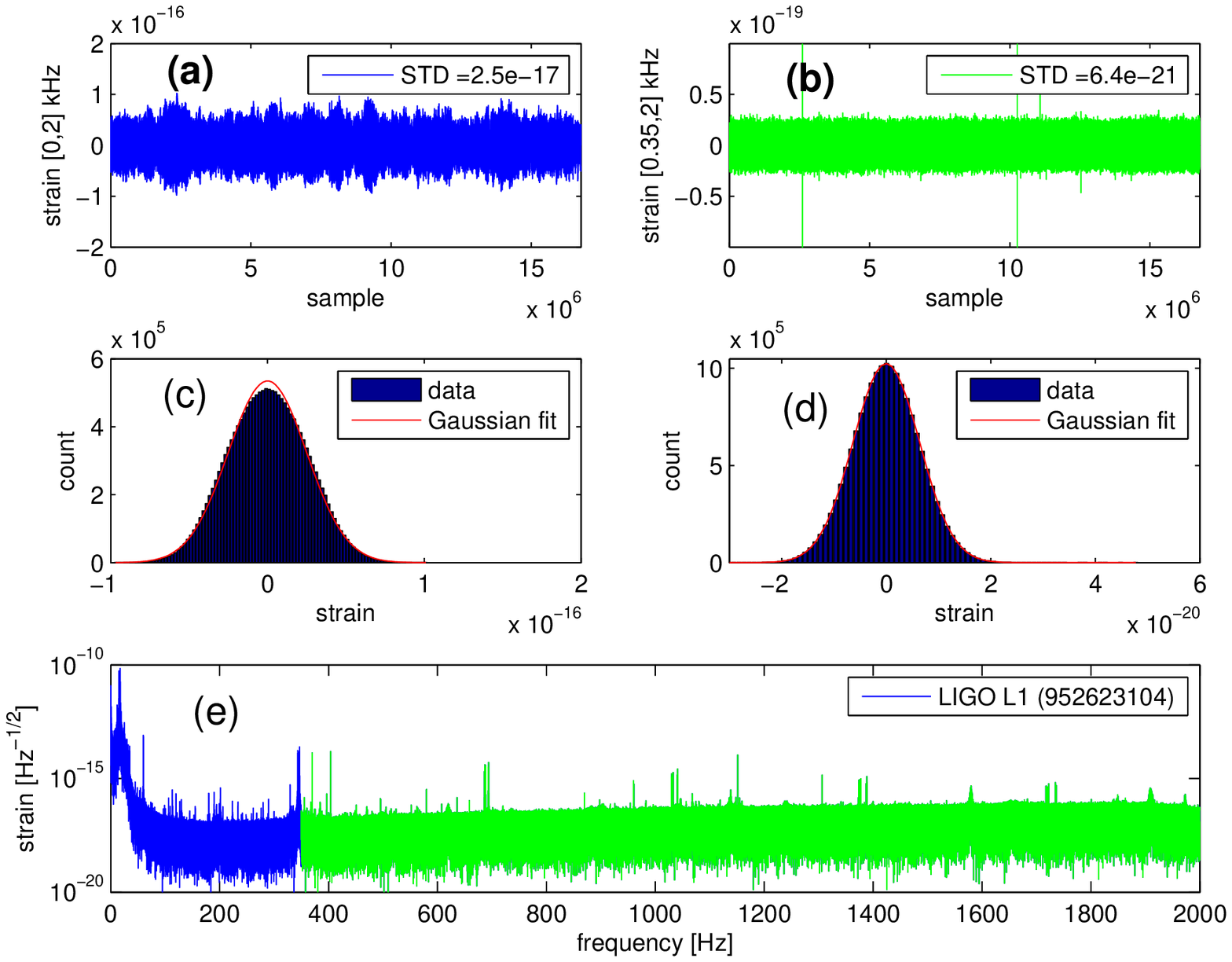}\includegraphics[scale=0.3]{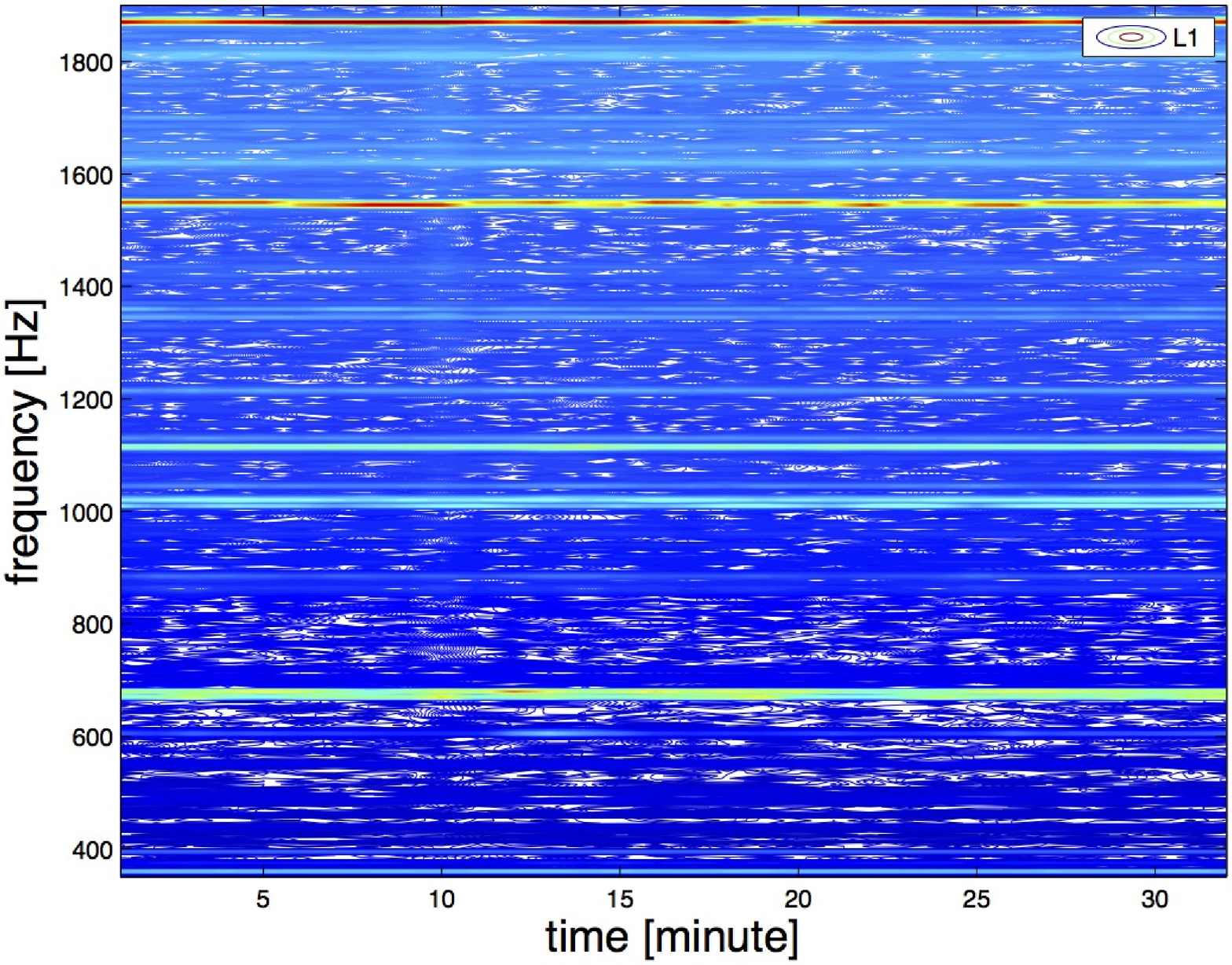}} 
\caption{(Left panels.) A high pass filter (350-2000 Hz) applied to LIGO S6 L1 data (0-4000Hz) in (a) shows the high frequency shot-noise dominated output in (b), here applied to 64 LIGO frames of 64 seconds each (4096$^2$ samples). Shot-noise is close to Gaussian (d), more so than the original broadband data (c). (Right panel.) Spectrogram of one-minute Fourier transforms over the first 32 seconds, marked by the presence of various lines.}
\label{fig1}
\end{figure}

We set out to probe for BEGE in SN 2010br by application of Time Sliced Matched Filtering (TSMF),
previously developed for analysis of noisy time-series of TAMA 300 and {\em BeppoSAX} \citep{van14}.
For high density matter at and about the ISCO around a stellar mass black hole as a source of gravitational waves, we focus on high frequencies and apply a bandpass filter of 350-2000 Hz (Fig. \ref{fig1}). The quality of noise in 350-2000 Hz is markedly different from its low frequency counter part below 350 Hz. At high frequency, shot noise arising from finite photon counts is nearly white. It is also of much smaller amplitude than the seismic-dominated noise below 350 Hz. Measured by standard deviation STD of strain data from the LIGO detector at Louisiana (L1) and
Hanford (H1) as shown in Fig. \ref{fig1}, high frequency noise is about three orders of magnitude smaller than low frequency noise. 

In focusing on the central engine, we seek to identify gravitational wave emission that may permit complete calorimetry
on the explosion process in Type Ib/c events and their associated LGRBs, that should be intimately related to
the evolution of angular momentum in aforementioned accretion flows and rotating black holes. 
Our focus is hereby distinct from the (many) other channels of gravitational wave emissions associated with
the formation and spin down of neutrons stars and black hole formation following core-collapse thereof. 
Their gravitational wave signatures tend to be different and of relatively short duration \citep[e.g.][]{ree74,due04,lip83,fry02,lip09,ott09,fry11,cer13}, that fall outside the scope and intent of the present search algorithm.  

Recently, several search algorithms for chirp like behavior have been developed based on pattern recognition in Fourier-based spectrograms \citep{sut10,pre12,thr13,thr14,cou15,abb15,gos15}.
However, chirps inherently show spreading of energy in frequency space (Fig. \ref{fig:0} below), i.e., spreading
over pixels in such spectrograms. (Fourier-based analysis is optimal for {\em horizontal} segments
in $ft$-diagrams.) While allowing large data sets to be analyzed relatively fast, these methods
are sub-optimal for chirps, relative to the theoretical maximum permitted by matched filtering.

Our approach is different. We consider a two-dimensional analysis, comprising a range of frequencies and 
their time rate-of-change. We endeavor to recover maximal sensitivity conform matched filtering in the limit of large template banks of size $O(\tau \max f[0,T])^2$, where $\tau$ refers to an intermediate time scale of chirp
template duration in TSMF. In \cite{van14}, we demonstrate that such large template banks are sufficient to extract complex broadband turbulence from noisy long duration time series. The proposed approach, however, heavily relies on
modern supercomputing for realistic searches. 

In \S2, we first revisit existing evidence for LGRBs from rotating black holes and prospects for long duration chirps in gravitational waves. We next describe our new butterfly filter in the time-frequency domain using TSMF (\S3). TSMF realizes near-maximal sensitivity in the application to noisy time series such as gamma-ray light curves of the {\em BeppoSAX} catalog \citep{van14} and gravitational strain data \citep{van15b}. Illustrative for its extreme sensitivity to broadband signals is the identification of a broadband Kolmogorov spectrum up to 1 kHz (in the laboratory frame) in bright GRBs with a mean photon count of 1.26 photons per 0.5 ms bin. The chirp templates used are superpositions of ascending and descending chirps by superposition, thus removing any bias to slope in the slope $df/dt$ of a chirp. We here port this method to the noisy time series of gravitational wave strain data of LIGO S6. In \S4, detection criteria are defined specific for BEGE, applied to two weeks of S6 data prior to SN2010br. In \S5, we conclude with an outlook on systematic probes of nearby events provided by robotic optical surveys.

\section{Current evidence for LGRBs from rotating black holes}

\begin{table*}[h]
\textbf{Table 1.} {Observational evidence for long gamma-ray bursts from rotating black holes}
\centerline{
\begin{tabular}{llllll}
\hline
{\sc Instrument} &  {\sc Observation/Analysis} &  {\sc Result} \\ 
\hline
\hline
{\em Swift} 	&      LGRBs with no SN, SGRBEE  				 & Mergers' Extended Emission [1] \\
				&      Amati spectral-energy correlation  		 & Universal to LGRB and SGRBEE [1]\\
				&      Discovery X-ray afterglow SGRBs		& SGRB 050509B [2]\\
{HETE-II} 			&      Discovery X-ray afterglow SGRBs		& SGRB 050709 [3] \\
 {\em BeppoSAX} 	&	Discovery X-ray afterglows LGRBs		& Common to LGRB and SGRB(EE) [1]\\	
 				&	Broadband Kolmogorov spectrum		& No signature of (proto-)pulsars  [4] \\	
 {\sf BATSE} 	&	$T_{90}^{SGRB}\simeq2$\,s, $T_{90}^{LGRB}\simeq20$\,s  & Short hard, long soft GRBs [5]\\
				&      ms time-variability					& Compact relativistic engines [6]\\
		 		&      Normalized light curves LGRBs	   		  & BH spin down against ISCO [7]\\ 	
{\sf Optical} 		&	LGRB association to SNIb/c			 & Branching ratio $<1\%$  [8]\\
 				&	Calorimetry SN kinetic energies		 & $E_{rot}>E_c$ in some GRB-SNe  [9]\\
\hline
\hline
\end{tabular}}
\label{TABLE1}
\small
1.\,Revisited in \cite{van14b}; 2.\,\cite{geh05}; 3.\,\cite{vil05,fox05,hjo05}; 
4.\,\cite{van14}; 5.\,\cite{kov93}; 6.\,\cite{sar97,pir97,kob97} (see also \cite{kob97,van00,nak02});
7.\,\cite{van09,van12,nat15}, see further \cite{van08b,sha15}; 8.\,\cite{van04,gue07}; 
9.\, In the model of \cite{bis70}, the rotational energy $E_{res}$ exceeds the maximal spin energy $E_c$ of a (proto-)neutron star in some hyper-energetic events \citep{van11a}.
\mbox{}\\\hskip0.08in
\end{table*}

Originating in SNIb/c, normal LGRBs derive from either black holes or neutron stars, the latter in the form of hot proto-neutron stars with possibly superstrong magnetized fields \citep{tho94,uso94,met11,pir11,pir12}. At a confidence level of greater than $4\sigma$, SGRBs originate in mergers \citep[][]{van14}. SGRBs should hereby be associated with black holes and especially so for the long-lasting soft Extended Emission (EE) in the {\em Swift} class of SGRBEEs;  the final outcome of a merger of a neutron star with another neutron star or companion black hole is always the same: a stellar mass black hole with practically the mass of the binary progenitor. The black hole from the former should be relatively rapidly rotating \citep[e.g.][]{bai08,van13}, whereas from the latter, its spin is not expected to be very different from its original spin prior to the merger and may be diverse \citep{van99}. As mentioned, the
phenomenology of SNIb/c's points to an explosion mechanism powered by a compact
angular momentum-rich inner engine.

On this basis, evidence for LGRBs from rotating black holes  derives from their association with SNIb/c and SGRB(EE)s by the following (Table 1):

\begin{enumerate}
\item {\em Universality.} Black hole inner engines to LGRBs point to X-ray afterglows also to SGRBs
\citep{van01}, confirmed by weak X-ray afterglows in the {\em Swift} event GRB050509B \citep{geh05} and the HETE-II event GRB050709 \citep{vil05,fox05,hjo05}. As a common inner engine, they also explain the observed extension of the Amati relation $E_{p,i} \propto E_{iso}^\alpha$ for LGRBs between the energy $E_{p,i}$ at the maximum of the $\nu F_\nu$ spectrum in the rest frame of the source and the isotropic-equivalent energies $E_{iso}$ \citep{ama02,ama06},
where $\alpha\simeq 0.5$, to the soft EE in the {\em Swift} class of SGRBEEs \citep{van14b}.
The BATSE durations $T_{90}\simeq 2$\,s and $T_{90}\simeq 20$\,s of SGRBs and, respectively, LGRBs can be explained by hyper- and suspended accretion states onto slowly, respectively, rapidly rotating black holes \citep{van01}.

\item {\em Long durations.} The proposed feedback of rapidly rotating black holes onto
matter at the ISCO \citep{van99,van03} gives canonical timescales of tens of seconds for the lifetime $T_{spin}$ of
initially rapidly rotating black holes, i.e., $T_{spin}$ is consistent with the observed $T_{90}$ durations in the BATSE catalogue. A model predicted correlation $E_{\gamma} \propto E_{p,i}T_{90}^{\beta}$ for the true energy in gamma rays $E_\gamma$ (corrected for beaming) is found in the data with $T_{90}\simeq T_{spin}$ from {\em Swift} and HETE-II  \citep{van08b} and BATSE \citep{sha15}, where $\beta\simeq 0.5.$ The time evolution of feedback has been tested against normalized light curves extracted from the 1493 LGRBs in the BATSE catalogue, showing fits to model light curves from black holes loosing angular momentum against high-density matter at the ISCO. The fits are especially tight for very long duration events $(T_{90}> 20$\,s), and more so than for feedback on matter further out or for model light curves from spin down of rapidly rotating neutron stars
\citep{van09,van12}. The same mechanism can account for the anomalously long durations of EE in SGRBEEs, which defy any dynamical time scale in mergers, upon associating SGRBEEs with mergers involving
rapidly rotating black holes. The latter naturally derive from binary mergers of neutron stars with
another neutron star or a rapidly rotating black hole companion. 

\item {\em Ample energy reservoir.} Calorimetry on the kinetic energies of supernovae 
associated with LGRBs reveals a some hyper-energetic events. In the scenario of \cite{bis70}, 
a few require rotational energies $E_{rot}$ in their inner engines well in excess of the 
maximal spin energy $E_c< 10^{53}$ erg a rapidly rotating (proto-)neutron star \citep{van11}. 
Rapidly rotating neutron stars or their magnetar variety are hereby ruled out as universal inner engines to LGRBs,
and most certainly so as universal inner engines to both LGRBs and SGRB(EE)s. Instead, the spin energy
$E_{spin}=3\times 10^{54} \,\mbox{erg}\left({M}/{10M_\odot}\right)\left[\sin(\lambda/4)/\sin(\pi/8)\right]^2$ $\left(-\frac{\pi}{2}\le\lambda<\frac{\pi}{2}\right)$ of rapidly rotating black holes can accommodate the most extreme  events,
even at moderate efficiencies.
 
 \item {\em Intermittency} in the prompt GRB light curves shows a positive correlation with brightness
 \citep{rei01}. (A few of the smoothest less-luminous events show Fast Rise Exponential 
 Decay (FRED) gamma-ray light curves \citep{rei01,van09}.) On the shortest time scales, intermittency
 has been identified with the inner engine, rather than processes downstream of the ultra-relativistic
 outflows powering the observed gamma-rays \citep{sar97,pir97}. Extreme luminosities 
 naturally derive from intermittent accretion onto the putative black hole \citep{van15a}. 
 In the presence of, e.g., violent instabilities in the inner accretion disk or torus, 
 the mass accumulated about the ISCO will be variable, resulting in on- and off-states
 described by a duty cycle $\tau/T$ given by the ratio of the durations $\tau$ of the on-state and 
a recurrence time $T$. The resulting mean luminosity of output in intermittently launched 
 magnetic winds scales with the inverse of the duty-cycle: $<L_w^i>  \propto  {T}/{\tau}$.

 \item {\em No signature of (proto-)pulsars.} A broadband Kolmogorov spectrum has recently been extracted
 from the 2kHz {\em BeppoSAX} light curves by high resolution matched filtering using 8.64 million
 chirp templates. The results reveal a smooth extension up to a few kHz in the comoving frame of 
 relatively bright LGRBs \citep{van14}. There is no ``bump" about a few kHz, as might be expected 
 from (proto-)pulsars. 
\end{enumerate}

\section{Gravitational waves from orbital and spin angular momentum}  

In light of the above, we here consider accretion flows onto initially near-extremal black holes 
as the putative inner engine to LGRBs \citep{van15b} and, by extension, a fraction of the progenitor Type Ib/c supernovae. Various aspects point to gravitational wave emission from non-axisymmetric mass motion.

Currently established experimental results are summarized in Table 2.
Decade long radio observations tracking the evolution of the Hulse-Taylor binary neutron star system 
shows orbital decay by gravitational wave emission in accord with the linearized
equations of general relativity to within 1\%. These gravitational wave emissions include the quadruple and 
various higher harmonics arising from the strongly elliptical orbit \citep{pet63}. This observational result 
establishes gravitational wave emission produced by multipole mass emissions in rotating systems with an ample energy 
reservoir in angular momentum. It demonstrates gravitational wave emission from a rotating, to leading order 
Newtonian tidal field. Mathematically, the latter acts as a source term to the linearized hyperbolic part of the
Einstein equations \citep[e.g.][]{van96}. As such, this mechanism is completely general and its applications should extend to rotating tidal fields in any other self-gravitating system, such as non-axisymmetric accretion flows in core-collapse 
events. The particular outcome will depend on the prospects for generating non-axisymmetric instabilities 
and the source of angular momentum driving the gravitational wave emission, i.e., {\em orbital angular momentum},
leading to contraction as in the Hulse-Taylor binary system, or {\em spin angular momentum}, leading to relaxation towards an approximately Schwarzschild space-time of slowly rotating central engines.

\begin{table*}[h]
\textbf{Table 2.} {Current experimental results on general relativity and high density matter.}
\centerline{\begin{tabular}{llllll}
\hline
{\sc Instrument} &  {\sc Observation/Analysis} &  {\sc Result} \\ 
\hline\hline
{\sf Radio}			&     Orbital decay NS binaries		& Multipole GW emission  [1,2]\\
{\sf LAGEOS-II}		&      Frame dragging (orbital)		& Asymptotic Kerr metric [3]\\
{\sf Gravity Probe B} &      Frame dragging (local) 			& Asymptotic Kerr metric [4]\\
{\sf Kamiokande, IMB} & 	$> 10$ MeV-neutrinos SN1987A	& Formation HD matter [5]\\ 
\hline
\hline
\end{tabular}}
\label{TABLE2}
\small \\
1.\,\cite{tay89,tay94,wei10}; 2.\,\cite{pet63}; 3.\,\cite{ciu04,ciu07,ciu09}; 4.\,\cite{eve11}; 5.\,\cite{bur87}.
\mbox{}\\\hskip0.08in
\end{table*}

\subsection{Type I: Ascending chirps}

Non-axisymmetric accretion flows have been widely considered to explain the observed Quasi-Periodic
Oscillations (QPO) in X-ray binaries or flaring in SgrA*, e.g., by magnetic stresses \citep{tag90,tag99,tag01,tag06,tag06b,lov14}, may have their counterparts in core-collapse events
with potential relevance to gravitational wave emission \citep{kob03,pir07,gam01,ric05,mej05,lov14,had14}),
including close to the ISCO stimulated by enhanced pressure, by heating or magnetic fields due to feedback 
by a rotating black hole \citep{van03,bro06}. Accretion onto the black hole may further excite quasi-normal mode ringing of the event horizon \citep[e.g.][]{ara04}. For stellar mass black holes produced in CC-SNe, however, their frequencies tend to be above the sensitivity bandwidth of ground based detectors  LIGO-Virgo and KAGRA.

In the extended accretion disk, cooling is perhaps most important in driving the formation of non-axisymmetric 
waves and structures. Self-gravity may hereby lead to fragmentation that, upon infall, will produce ascending
chirps \citep{pir07}. For a recent detailed discussion on these ascending chirps, see, e.g., \cite{gos15}. 
Alternatively, non-axisymmetric wave patterns of sufficient amplitude may become 
sufficiently luminous in gravitational waves, such that their radiation in angular momentum dominants over
viscous angular momentum loss. If so, these wave patterns likewise produce ascending chirps 
\citep{lev15}. These alternatives serve to illustrate various generic and potentially natural conditions for
ascending chirps to derive from non-axisymmetric accretion flows, converting orbital angular momentum
into gravitational waves.

In \cite{lev15}, we discuss a detailed framework for gravitational wave emission from wave patterns in accretion disks (Fig. 3). A primary control parameter is the efficiency in angular momentum loss by quadrupole gravitational radiation over viscosity mediated transport, parameterized by 
\begin{eqnarray}
P = \frac{\xi}{\alpha},
\label{EQN_P}
\end{eqnarray}
where $\xi$ is a dimensionless amplitude of a quadrupole mass-inhomogeneity and $\alpha$ refers to 
the $\alpha$-disk model, parameterizing the kinematic viscosity $\nu=\alpha c_s H$ in terms of the isothermal
sound speed $c_s$ and the vertical scale height $H(r)=\eta r$ of the disk. In what follows,
we shall write $\alpha=0.1\alpha_{-1}$, $\eta=0.1\eta_{-1}$, $P=10P_1$ in the inner region about the black hole 
of mass $M=10M_1\,M_\odot$ at a distance $D=D_2\,100\,\mbox{Mpc}$. The resulting spectrum of 
the dimensionless characteristic strain amplitude $h_{char}(f)$
is parameterized by a break set by the critical radius $r_b$ (if greater than the ISCO radius $r_{ISCO}$), within 
which gravitational radiation losses dominates over outwards viscous 
transport in angular momentum. Fig. 1 shows the characteristic {break} frequency $f_b$ 
in the turn-over of the green curves, satisfyng
\begin{equation}
f_b=430\,\eta_{-1}^{9/2}M_1^{-1}P_{-1}^{-3} \left( \frac{\dot{M}}{M_\odot\mbox{s}^{-1}}\right)^{-3/2}\,{\rm Hz}.
\label{f_b}
\end{equation}
The condition $r_b>r_{ISCO}$ implies $f_b<2f_{ISCO}$, where for a $10 M_\odot$ black hole $430\ {\rm Hz}<2f_{ISCO}<3000\ {\rm Hz}$, depending on the spin parameter $a/M$ of the black hole,  with $2f_{ISCO}=1600$ Hz for $a/M=0.95$ as an example.
By the orientation-averaged characteristic strain amplitude \citep{fla98,cut02}
\begin{eqnarray}
h_{char}\left(f\right) = \frac{\sqrt{2}}{\pi D} \sqrt{\left| \frac{\Delta E}{\Delta f} \right| },
\label{EQN_hchar}
\end{eqnarray}
where $\Delta E/\Delta f$ is the one-sided spectral-energy density at gravitational wave  frequency $f$,
we have
\begin{equation}
h_{char}(f)=\kappa \left\{\begin{array}{ll}
          \left(\frac{f}{f_b}\right)^{1/6}
 & \mbox{($f <f_b$)}\\
        \left(\frac{f}{f_b}\right)^{-1/6}\sqrt{2-\left(\frac{f_b}{f}\right)^{2/3}} & \mbox{($f_{isco}>f > f_b$)}\end{array} \right. 
\label{h_char-tot}
\end{equation}
with   
\begin{equation}
\kappa=\sqrt{8}\times10^{-22}D_{2}^{-1} M_1^{1/3}\left(\frac{\dot{M}\tau}{0.1M_\odot}\right)^{1/2}
\left(\frac{f_b}{1000  Hz}\right)^{-1/6}
\label{kappa}
\end{equation}
in terms of the mass accretion rate $\dot{M}$  and
a lifetime $\tau$ of the disk pattern. 
(\ref{h_char-tot}) applies generally, also to $f_b>f_{ISCO}$ for which $h_{char}=\kappa (f/f_b)^{1/6}$
in the range of $f$ less or equal that the quadrupole gravitational wave frequency at the ISCO.
  
\subsection{Type II: Descending chirps}

The gravitational wave luminosity of a mass moment $I_{lm}$ in at torus about the ISCO,
expressed in the quantum numbers $l$ and $m$ of spherical harmonics, satisfies
\citep{tho80,bro06}
\begin{eqnarray}
L_{GW} \propto \Omega_T^{2l+2}I_{lm}^2,
\label{EQN_LQ1}
\end{eqnarray}
whereby $m=l$ is the most luminous \cite{bro06}. To leading order $\Omega_T^2\propto r^{-3}$, whereby
(\ref{EQN_LQ1}) satisfies
\begin{eqnarray}
L_{GW} \propto \frac{1}{r^{m+3}}.
\label{EQN_Q2}
\end{eqnarray}
Gravitational wave luminosity hereby tends to reach maximum at the ISCO. 
We here attribute $I_{lm}$ to instabilities induced by enhanced thermal and magnetic pressures induced by feedback from the black hole \citep{van02,bro06}, provided it spins rapidly $(\Omega_H > \Omega_T)$. 
To be more precise, the feedback is envisioned to exceeds the luminosity in magnetic winds with $\Omega_H/\Omega_{ISCO}>1$ (up to 1.4396 according to the Kerr metric). For instance, if the \cite{sha73} solution serves as a leading order approximation of the inner disk, the total energy output $E_H$ onto the ISCO exceeds energy losses $E_w^*$ in magnetic winds whenever
  $a/M\ge 0.4433$ \citep{van12}. Equivalently, the initial rotational energy merely exceeds the 9\% of the maximal rotational energy by about 9\%, which is a rather mild condition. The excess $E_H - E_w^*>0$ is available to gravitational waves and MeV neutrinos. 
The gravitational wave spectra from these instabilities tends to be dominated by the lowest order multipole mass moments. Consequently, {\em a dominant output in gravitational waves is expected in quadrupole emission from the ISCO.} 
Consider the quadrupole emission formula \citep{pet63}, in geometrical units
\begin{eqnarray}
L_{GW} = \frac{32}{5} \left(\mu\Omega\right)^\frac{10}{3}
\label{EQN_Q3}
\end{eqnarray}
scaled to $L_0=c^5/G=3.64\times 10^{59}$ erg\,s$^{-1}$, where $\Omega$ denotes the orbital angular velocity
and $\mu$ the chirp mass. Expanding into small mass perturbations $\delta m$ and ignoring grey body
factors that may arise from proximity to the black hole,  for a dimensionless inhomogeneity $\xi=\delta m/M_T$ 
and a torus of mass $\sigma=M_T/M$ at $r_{ISCO}$ around a black hole of mass $M$, (\ref{EQN_Q3}) reduces to
\begin{eqnarray}
L_{GW} = 2\times 10^{51} \left(\frac{\xi}{0.1}\right)^2 \left(\frac{\sigma}{0.01}\right)^2\left(\frac{4M}{r_{ISCO}}\right)^{5}~\mbox{erg~s}^{-1}.
\label{EQN_Q4}
\end{eqnarray} 
The observed instantaneous dimensionless strain at a source distance $D$ satisfies
\begin{eqnarray}
h=0.7\times 10^{-23}\,M_1D_2^{-1} \left(\frac{\xi}{0.1}\right)\,\left(\frac{\sigma}{0.01}\right)
\left(\frac{f}{600\,\mbox{Hz}}\right)^{\frac{2}{3}},
\label{EQN_Q5}
\end{eqnarray}
where $f=2f_{orb}$ in the Newtonian approximation $2 \pi f_{orb} = M^{-1} (M/a)^{3/2}$ and $M=M_1\,10M_\odot$.

Mediated by relativistic frame-dragging, rotating black holes may sustain (\ref{EQN_Q4}) for extended durations
by sustained feedback onto matter at the ISCO via an inner torus magnetosphere \citep{van99}. Non-relativistic frame-dragging has recently been experimentally established by LAGEOS-II and Gravity Probe B. Feedback will be particularly 
prominent in the presence of intermittencies \citep{van15a}, e.g., arising from various 
instabilities in accretion flows and the feedback mechanism itself through an inner torus magnetosphere. 
In deriving (\ref{EQN_Q4}) from angular momentum in spin of the black hole, the latter will gradually
slow down with an accompanying expansion of the ISCO. The result is a descending chirp. 

Exact solutions to such descending chirps derive from solutions to feedback in the Kerr metric 
\citep{van08,van12}, defined by the initial value problem of conservation of total energy and angular
momentum 
\begin{eqnarray}
\dot{M}  = \Omega_T \dot{J},~~\dot{J}  =-\kappa e_k (\Omega_H-\Omega_T),
\label{EQN_ode1}
\end{eqnarray}
where $\Omega_H$ and $\Omega_T$ denote the angular momentum of the black hole and, respectively, a torus in
suspended accretion at the ISCO. 
Normalized light curves of LGRBs in BATSE point to initially near-extremal black holes \citep{van15b}. Important to rapidly rotating black holes, is its lowest energy state that preserves maximal horizon flux by an equilibrium value of Carter's magnetic moment \citep{car68}. Thus, $\kappa\simeq uM$ describes the strength of the feedback, parameterized by the ratio $u\simeq 1/15$ total energy ${E}_{B,p}$ in poloidal magnetic field relative to the kinetic energy ${ E}_k$ of a torus at the ISCO \citep{van03}. The energy per unit mass satisfies$e_k\simeq\frac{1}{2}v^2 e$, where $v=Mz\Omega_{ISCO}$ and $e$ denotes the specific energy at the ISCO in the Kerr metric \citep{bar72}. 

In re-radiation of input received by (\ref{EQN_ode1}), matter at the ISCO effectively acts as a catalytic converter of
the rotational energy of the black hole, wherein the amplitude $\xi$ in (\ref{EQN_Q4}-\ref{EQN_Q5}) is determined
self-consistently \citep{van12}.  Numerical integration of (\ref{EQN_ode1}) hereby defines descending chirps based on the evolution of the orbital frequency $f_{ISCO}(t)$ at the ISCO as a function of time $t$. Specifically, multipole mass moments at the ISCO radiate at multiples $mf_{ISCO}$ $(m=2,3,\cdots)$, e.g., from ISCO waves \citep{van02}. 

\begin{figure}[h]
\centerline{\includegraphics[scale=0.45]{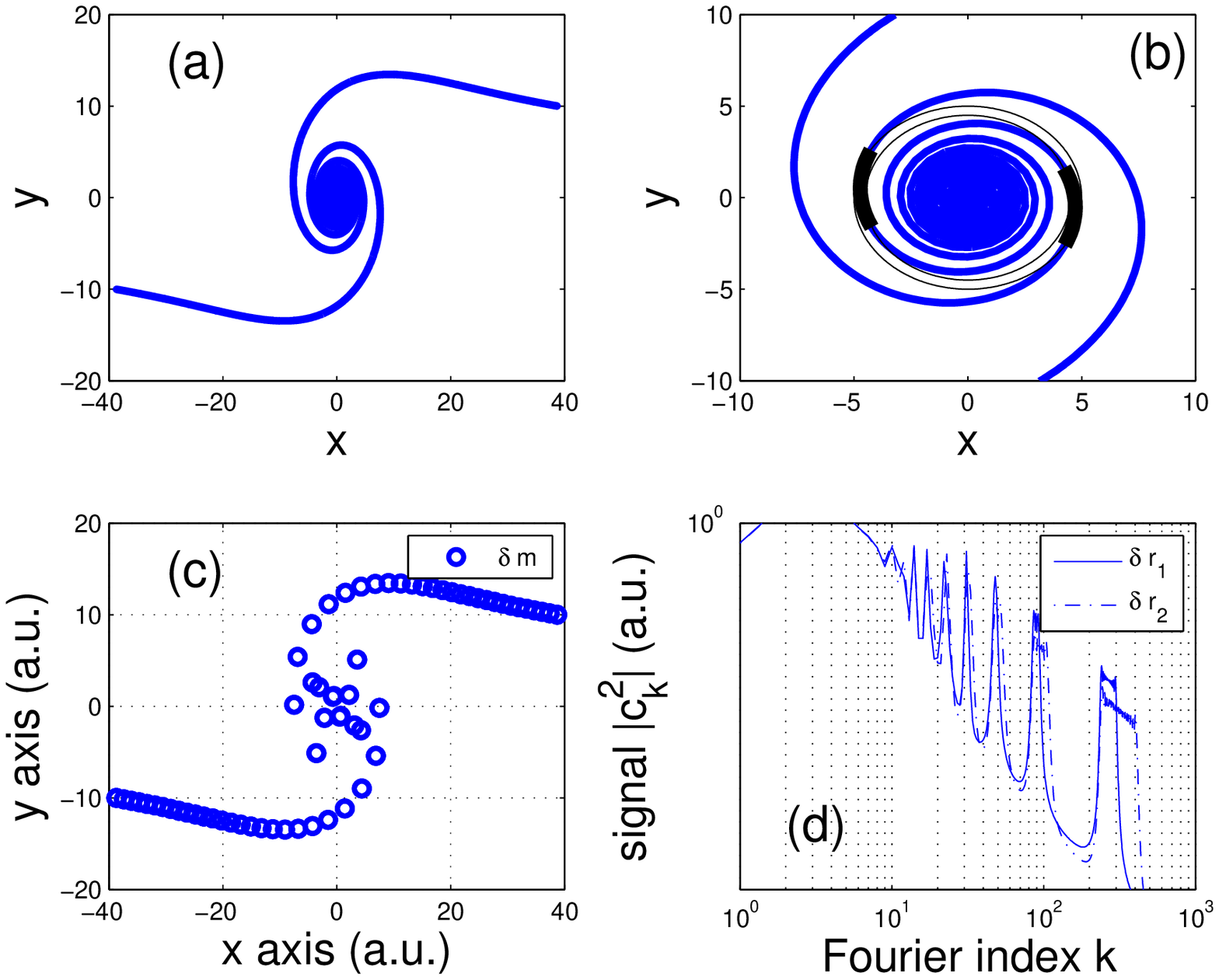} \includegraphics[scale=0.45]{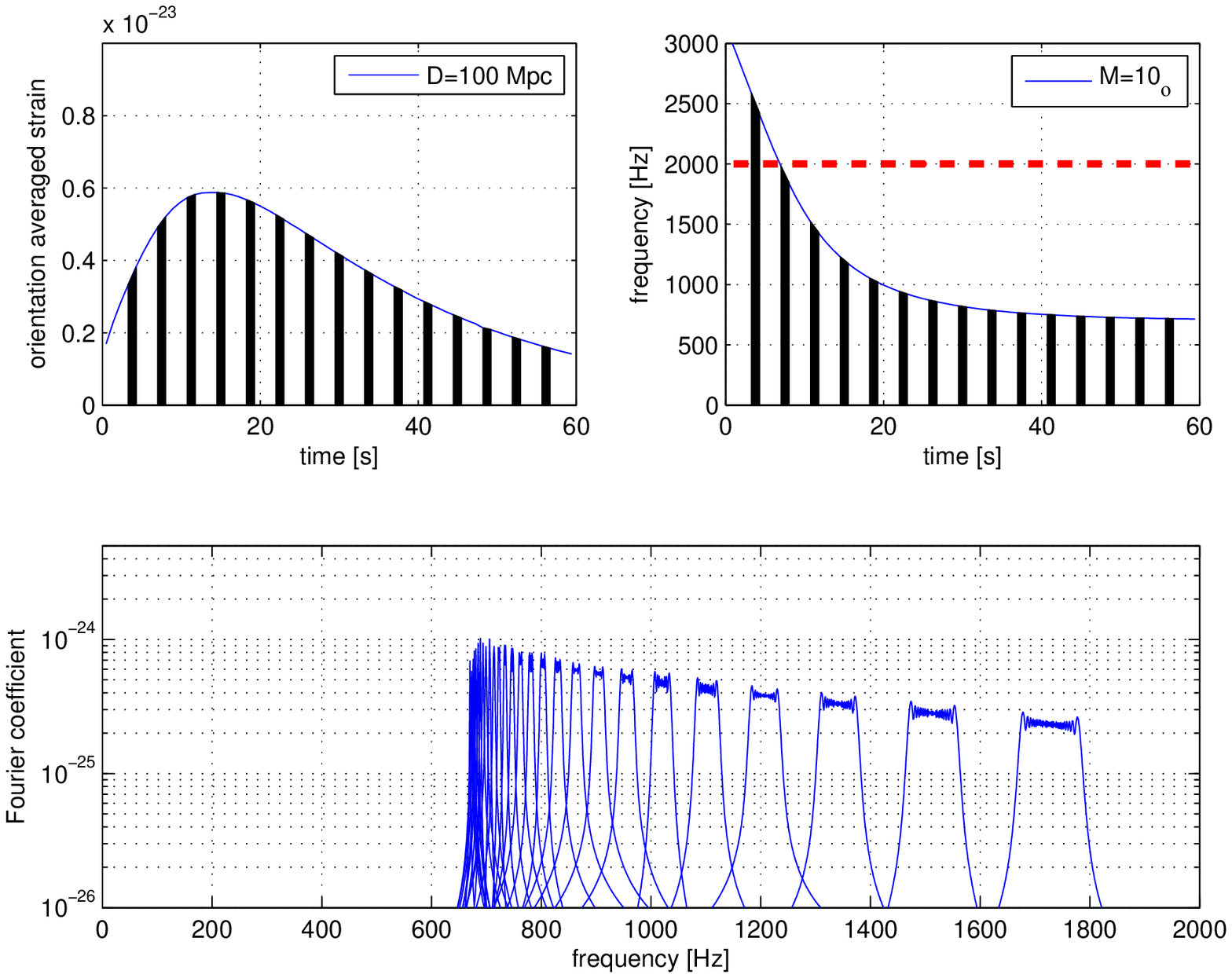}}
\caption{(Left.) (a) Shown is a spiral density wave pattern in an accretion disk. (b) Over-dense regions (thick black) have finite angular extend in an annular region $4.5 < r < 5$ (thin black circles). Approximating the latter by local mass-inhomogeneities $\delta m$ in (c) at Keplerian rotation implies quadrupole emission spectra of each $\delta m$, that 
are non-overlapping in frequency (d), here expressed by Fourier coefficients $|c_k|^2$. Spectral broadening due to accretion is shown for two accretion rates corresponding to radial migrations $\delta r_2<\delta r_1<0$. (Reprinted from \cite{lev15}.) (Right.) Theoretical wave form of ISCO waves due to feedback by an initially extremal black hole, defined by the initial value problem (\ref{EQN_ode1}). Shown is the orientation averaged strain amplitude $h(t)/\sqrt{5}$. In a turbulent background accretion flow, phase-incoherence is anticipated to be limited to an intermediate time scale $\tau$, requiring time-slicing in the application of matched filtering, here shown for $\tau=1$ s. The red dashed line
refers to the high frequency cut-off in the 4096 Hz down sampled LIGO data. (Adapted from from \cite{van12b}.)}
\label{fig:0}
\end{figure}

According to (\ref{EQN_ode1}), solutions are determined by two fixed points, namely, when the angular velocity of the black hole and that of matter at the ISCO are the same: initially at $t=0$ and at late times as $t$. The first features an initial strengthening in the feedback process, the latter an essentially exponential decay in frequency with time. By the second fixed point, the gravitational wave frequency $f(t)$ as a function of time and total duration $T$ is well approximated by an overall exponential decay in frequency,
\begin{eqnarray}
f_{ISCO}(t) = f_1 + \left(f_0-f_1\right)e^{-at/T}
\label{EQN_ft}
\end{eqnarray}
where $a$ is a dimensionless scale. The late-time asymptotic frequency of a black hole of
initial mass $M$ satisfies a limited range \citep{van11}
\begin{eqnarray}
595\,\mbox{Hz} \left(\frac{10M_\odot}{M}\right)
\le f_{GW} \le 704\,\mbox{Hz}\left(\frac{10M_\odot}{M}\right)
\end{eqnarray}
for $f_{GW}=2f_{ISCO}$ by a finite dependency of the total energy output on the initial black hole spin. (The lowest frequency corresponds to an initially extremal black hole with maximal energy output.)
Asymptotic analysis shows a dominant emission in gravitational waves over any accompanying output in jets, MeV-neutrinos and magnetic winds \citep{van03,van15b}. 
The total energy output typically reaches fraction of order unity of the initial rotational energy of the black hole.

Fig. \ref{fig:0} illustrates the combined outlook of ascending and descending chirps. 

\section{Chirp search by a butterfly filter in time-frequency}

Characteristic for BEGE are (a) trajectories $f(t)$ in time-frequency space that are non-constant in frequency, i.e., with finite slopes $df/dt$ and (b) a probably loss of phase coherence over a large number of wave periods, due to its origin in (magneto-)hydrodynamical mass motion. This suggests searches for BEGE by applying a bandpass filter to $df/dt$, here in the form of finite bandwidths of chirp templates of intermediate duration $\tau=1$ s, for correlated with strain data by application of aforementioned TSMF. Our chirp templates are extracted by time slicing a long duration model chirp.

By using superpositions of ascending and descending chirps, our complex chirp templates have no bias to slope in the slope $df/dt$ of a chirp \citep{van14}.  When using banks of millions of templates, TSMF densely covers the frequency 
range and frequency time rate-of-change. This power of TSMF is demonstrated by identifying complex and
broadband Kolmogorov spectra in light curves of long GRBs up to 1 kHz (in the observer's frame of reference), 
extracted from BeppoSAX light curves with, on average, merely 1.26 photons in each bin of 500 $\mu$s. 

In the present application to LIGO time series, we partition LIGO data into frames of one minute, comprising 64 seconds of $n=2^{18}$ samples. At a down sampled frequency of 4096 Hz, the LOSC data give a sensitivity bandwidth of 0-2000 Hz. 

\begin{figure}[h]
\centerline{\includegraphics[width=140mm,height=80mm]{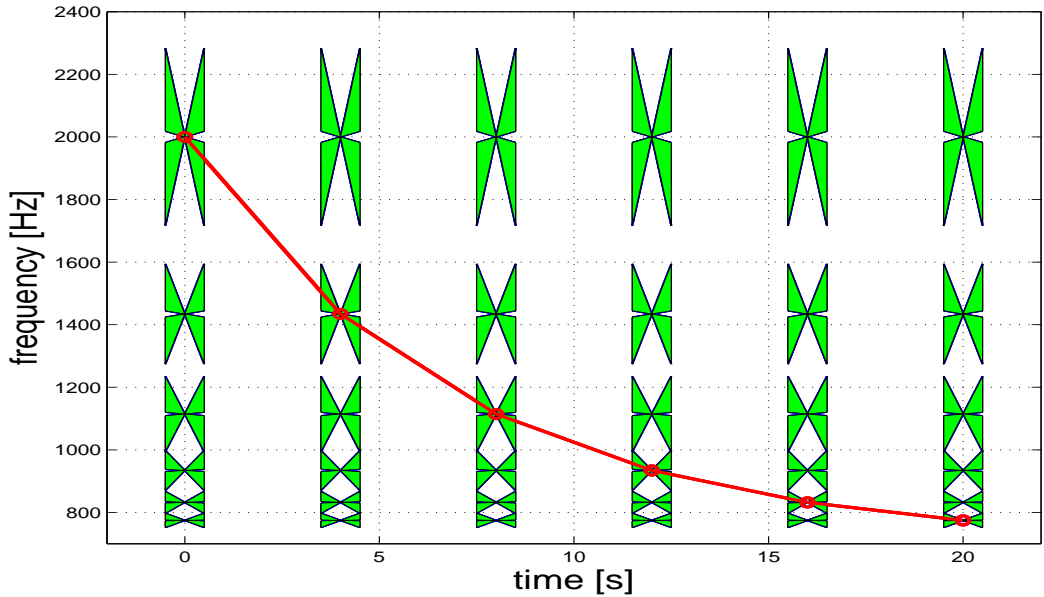}}
\caption{Shown is our butterfly filter in the time-frequency plane, to capture trajectories of long duration model chirp by TSMF, here in terms of chirp templates of intermediate duration $\tau=1$ s with varying bandwidth. The vertex and opening angle of each butterfly represent a central frequency $f$ and, respectively, a bandpass of slopes $df/dt$.}
\label{fig3a}
\end{figure}
\begin{figure}[h]
\centerline{
\includegraphics[width=82mm,height=60mm]{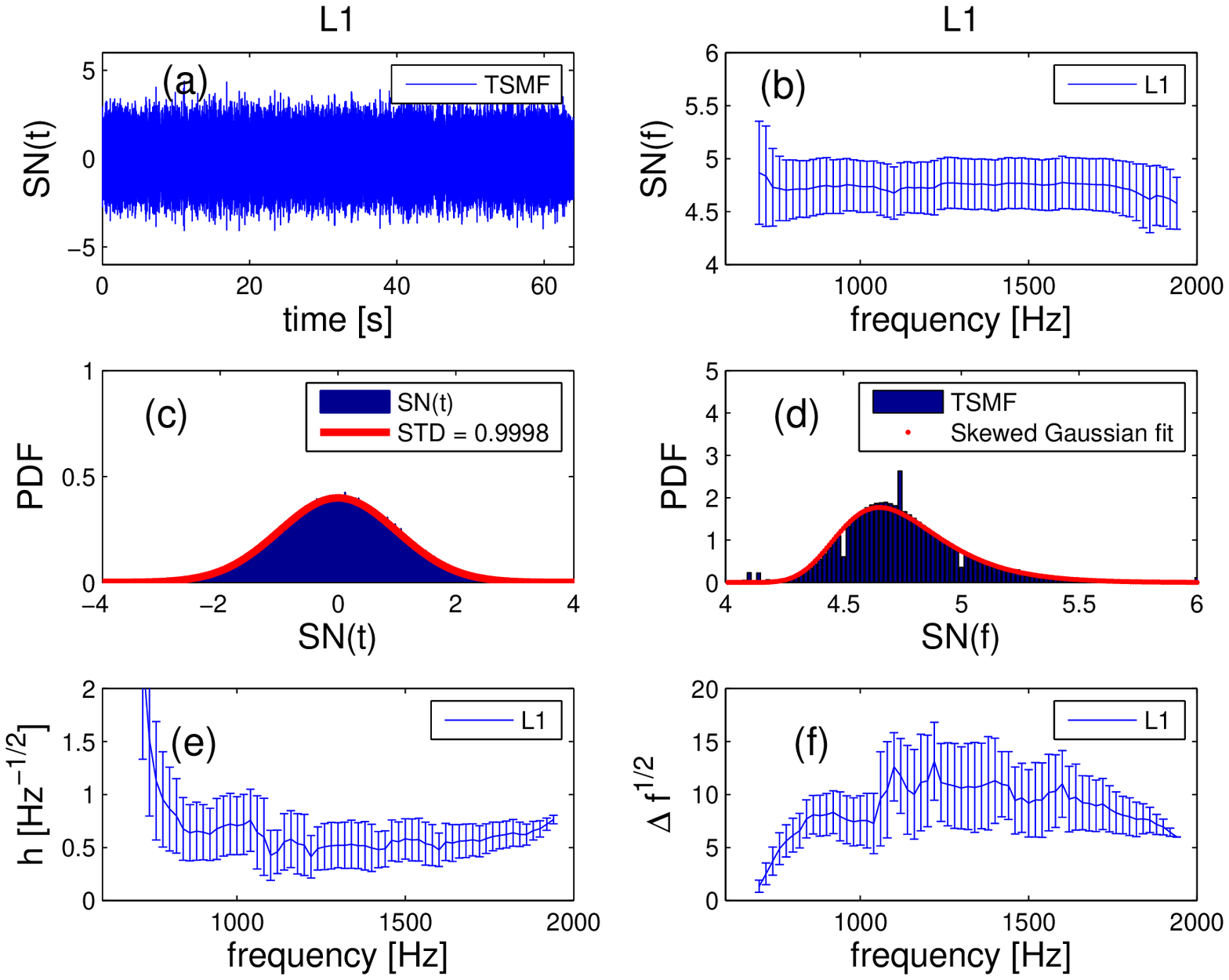}
\includegraphics[width=82mm,height=60mm]{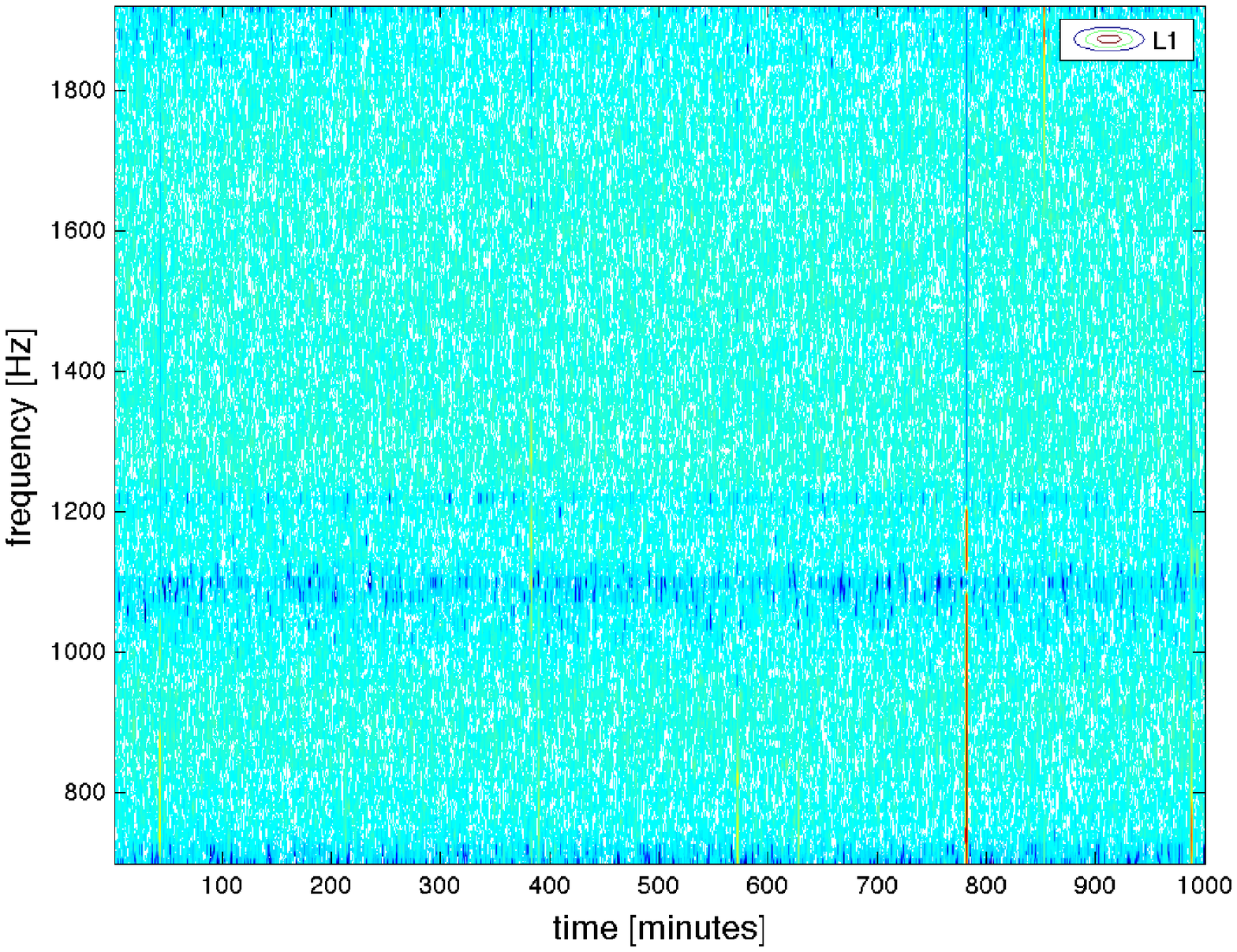}}
\caption{Overview of TSMF applied to LIGO S6 data of the L1 detector. (Left six panels.) (a) SN($t$) is defined by normalized Pearson coefficients as a function of time $t$ produced by a chirp template of $\tau=1$ s in TSMF. 
The maxima SN($f$) of SN($t$) as a function of frequency obtain a pseudo-spectrum, based on a bank of templates covering a broad range of chirp frequency $f$ and bandwidth $\Delta f$. (b) Shown is the mean $<$SN($f$)$>$ over all frames over a two week period including its 1 $\sigma$ variations. (c) SN($t$) satisfies a Gaussian distribution, here normalized to unit variance in the application to a one minute frame of LIGO data ($n=2^{18}$ samples). 
 (d) SN($f$) satisfy a skewed Gaussian PDF.  (e-f) The true spectrum is somewhat non-smooth, as our template bank covers a section of $(f,\Delta f)$ space somewhat non-uniformly with a small jump around 1100 Hz. 
(Right panel.) The spectrogram of SN($f)$ 
is essentially featureless with noticeable absence of any lines, suppressed by the butterflies shown in Fig. \ref{fig3a}, that impose a bandpass for nonzero slopes $df/dt$. The non-uniformity in our template bank is apparent in a weak
noisy feature about 1100 Hz.}
\label{fig4a}
\end{figure}

With $\tau=1$ s, each choice of model parameters $f_0$ and $T$ gives 64 chirp templates by time slicing,
having a mean frequency $f_i$ and bandwidth $B_i=\Delta f_i$, where $i=1,2,\cdots 64$. The distribution
of chirps represents a cover of the time-frequency plane $(t,f)$ with ``butterflies" $(f,df/dt)$, defined by
a frequency $f$ at its vertex and a range of slopes $df/dt$ defined by the template bank (Fig. \ref{fig3a}).
This approach enables capturing a wide variety of trajectories in the $(t,f)$-plane comprising ascending and descending
chirps. A recent demonstration of the power of this approach is the identification of broadband turbulence
in the noisy gamma-ray burst time series of long GRBs from the {\em BeppoSAX} catalog \citep{van14}.
Fourier transforms correspond to the degenerate case of butterflies with zero opening angle.

TSMF calculates correlations SN($t$) by convolving chirp templates with strain from normalized Pearson coefficients.
For chirp templates of intermediate duration $\tau=1$ second applied to LIGO frames of
$2^{18}$ samples, these correlations satisfy a truncated distribution that is close to Gaussian with unit variance, 
where the truncation is set by the correlation length $n$. These truncations satisfy a skewed Gaussian with PDF
(\ref{fig4a})
\begin{eqnarray}
p(y) = \frac{ n\,\mbox{erf}(y)^{n-1}}{\sqrt{2\pi}\sigma} e^{-y^2},
\label{EQN_SG}
\end{eqnarray}
where $y={x}/{\sqrt{2\pi}\sigma}$ for a standard deviation $\sigma$, derived from the probability $P(<x)=\int_0^x p(s)ds$ that all $n$ correlations are below $x$. 

For a given LIGO frame and model parameters, TSMF identifies, out of 64 one-second slices (4096 samples each), the chirp templates with maximal correlation SN($f$), where $f$ denotes the frequency of the associated chirp (Fig. \ref{fig3a}). Thus, SN($f)$ is essentially uniform in frequency (white), whose PDF is a skewed Gaussian (Fig.\ref{fig4a}). It should be emphasized that SN($f$) is the pseudo-spectrum, in contrast to a real spectrum defined by the average $<$SN($f$)$/\sqrt{\Delta f}>$ over chirps of frequency $f$ with bandwidth $\Delta f$ \citep{van14}.

To efficiently search for BEGE, we propose a two-step process in which a coarse grained search identifies
epochs of interest for fine grained searches using a large number of chirp templates. For the signals of
interest, the latter approaches the theoretical limit of sensitivity of matched filtering using up to ten 
million chirp templates \citep{van14,van15b}. A relatively coarse grained search using fewer templates
will enable an initial scan of epochs of long durations, e.g., weeks or more. Here, we defined coarse and fine grained searches by the number of scaled parameters in the long duration model chirps (\ref{EQN_ft}).
Preferred parameters are, for instance, overall durations $T$ and scaling of the frequencies
$f_0$ and $f_1$ in (\ref{EQN_ft}). Upon scaling of $T$ only, Fig. \ref{fig5} demonstrates the
sensitivity using 64 thousand chirp templates. It is sub-optimal by a factor of about 0.6 relative
to aforementioned fine grained searches.

To identify events of interest at SN$>6$, the coarse grained search shown in Fig. \ref{fig5} is suitable
to scan a large epoch such as weeks prior to a nearby core-collapse supernova.

\section{Pseudo-spectra SN($f)$ from TSMF are essentially optimal for monotonic chirps}

Matched filtering gives a theoretical upper bound for the effective dimensionless strain, satisfying \citep{van01a}
\begin{eqnarray}
h_{eff} \simeq \left(\frac{M_1}{D_{2}}\right) \left(\frac{E^{GW}_{-1}}{M_1}\right)^\frac{1}{2}\simeq 10^{-21},
\label{EQN_heff1}
\end{eqnarray}
where $D=100D_{2}$ Mpc, $M=10 M_1 \,M_\odot$ and $E^{GW}=0.1 E^{GW}_{-1} M_\odot$ denotes the energy
output in gravitational waves. The theoretical upper bound is based on a perfect match to a model template
over the full duration of the burst in the face of Gaussian noise. 
In TSMF, matched filtering is partitioned over intermediate time intervals $\tau$ of phase coherence, since
 no (predictable) phase coherence is expected over the full duration $T$ of the burst for the present
 (magneto-)hydrodynamical source under consideration. Results for single chirp templates of duration
 $\tau$ hereby define a partition over $N=T_{90}/\tau$ time slices, each satisfying \citep{van11,van15b}
\begin{eqnarray}
h_{eff}^\prime \simeq \frac{1}{\sqrt{N}} h_{eff} \sim 10^{-22}
\label{EQN_heff2}
\end{eqnarray}
when $T$ matches $T_{90}$ of the burst, here identified with the durations of tens of seconds of long GRBs.

\begin{figure}[h]
\centerline{\includegraphics[scale=0.51]{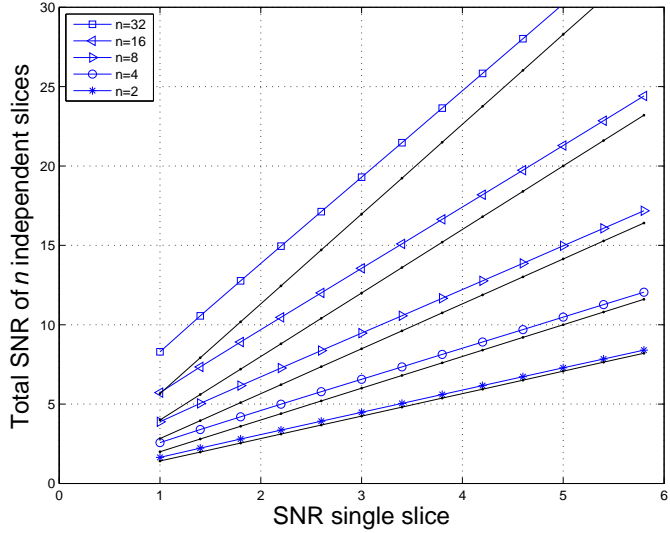}}
\caption{The approximate expression $\sqrt{N}\left<\sigma_i\right>$ (black lines) gives a conservative estimate for  the equivalent total level of confidence $\sigma_{tot}$ (blue tagged lines) in TSMF.}
\label{fig001}
\end{figure}
 
To combine results (\ref{EQN_heff2}) obtained from different slices, consider the partition of a time interval $[0,T]$ in intervals $I_i=[t_{i-1},t_i]$ $(t_i=(i-1)\Delta t$, $i=1,2,\cdots,N$), and slicing of a long duration model chirp $f(t)$ into $N$ chirp templates on the $I_i$. When $f(t)$ is strictly monotonic, these chirp templates
have non-overlapping frequency intervals $F_i=[f_{0i},f_{1i}]$, $f_{0i}=\min f[I],$ $f_{1i}=\max f(I_i)$. 
The joint probabilities over different slices are hereby joint probabilities over different frequency bins. Since
LIGO shot noise is essentially white (Fig. 5), matched filtering on the $I_i$ will be statistically 
independent. Joint probabilities over various $I_i$ hereby reduce to ordinary probability products. 
The equivalent $\sigma_{tot}$ to $N$ confidence levels $\sigma_i$ in candidate detectors on the $I_i$
thus derive from the associated probabilities $p_i$ as
\begin{eqnarray}
\sigma_{tot} = \sqrt{2}\,\mbox{erfc}^{-1}\left(\prod_{i=1}^N p_i\right),~~
p_i =\mbox{erfc}\left( \frac{\sigma_i}{\sqrt{2}}\right),~~
 \left< \sigma_i\right> = \frac{1}{{N}} \sum_{i=1}^N \sigma_i,
\end{eqnarray}
where the latter expresses the mean of the confidence levels $\sigma_i$ for each time slice. 
Numerical evaluation shows to good approximation (Fig. \ref{fig001})
\begin{eqnarray}
\sigma_{tot} \simeq \sqrt{N} \left< \sigma_i\right>
\end{eqnarray}
for the total confidence level. (The right hand side is  a lower bound on $\sigma_{tot}$.)
This result recovers the familiar scaling with the square root of the number of wave periods in
matched filtering applied over the original time interval $[0,T]$. 
In searches for strictly monotonic chirps, we consider the output of TSMF 
in terms of a pseudo-spectrum shown in Fig. \ref{fig4a}, 
expressed by confidence levels $\sigma_i$ over the adjacent frequencies $F_i$.
Consequently, {\em TMSF recovers the ideal sensitivity limit (\ref{EQN_heff1}) at full resolution in a bank of about 
$(\tau \max f[0,T])^2$ templates}. 

\begin{figure}[h]
\centerline{\includegraphics[scale=0.65]{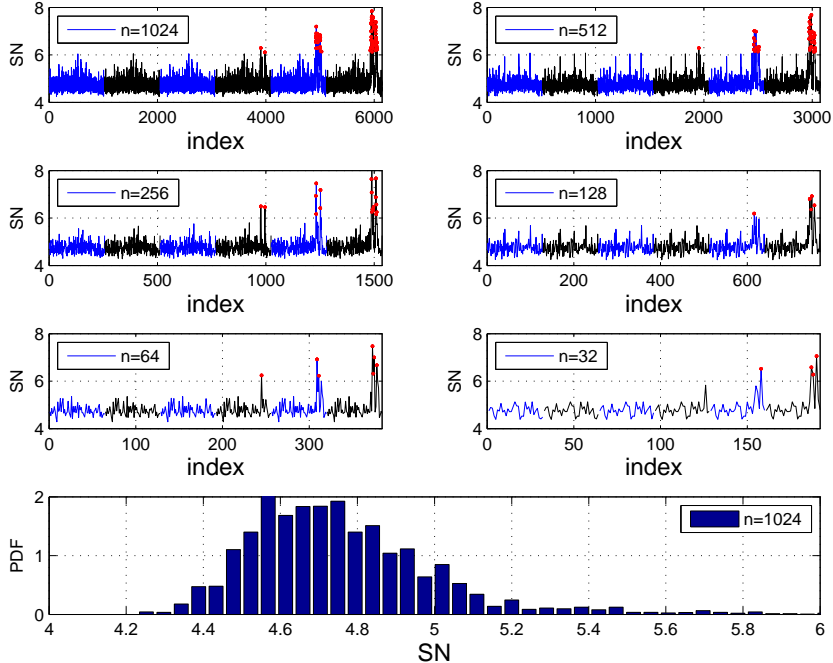}} 
\caption{(Top panels.) Sensitivity analysis on a test chirp of $T_s=20$ s duration decaying
exponentially down to 720 Hz from initially 2300 Hz, by injection into a one minute LIGO L1 frame analyzed by our butterfly TSMF algorithm. The alternatively blue and black segments refer to scans
with consecutively increasing signal strength, here signal-to-noise ratios expressed by ratios of
standard deviation ranging over six values $\{0, 0.0560,0.1121,0.1681,0.2242,0.2802\}$. The panels
show various degrees of coarse graining $n=1024,512,256,128,64,32$, 
here of model chirps of various durations of about 
$T=0.4-40$ s with slicing into chirp templates of $\tau=1$ s. Events of interest with SN$>6.1$ are
indicated in red. (Bottom panel.) Shown is the PDF of the
SN values for $n=1024$. The sensitivity achieved in this coarse grained search is about 60\% 
of the sensitivity achieved with a two-parameter search, by scaling of durations and
frequencies in model chirps (\ref{EQN_ft}).}
\label{fig5}
\end{figure}
 
\section{Detection criteria for BEGE}

\begin{figure}[h]
\centerline{\includegraphics[scale=0.46]{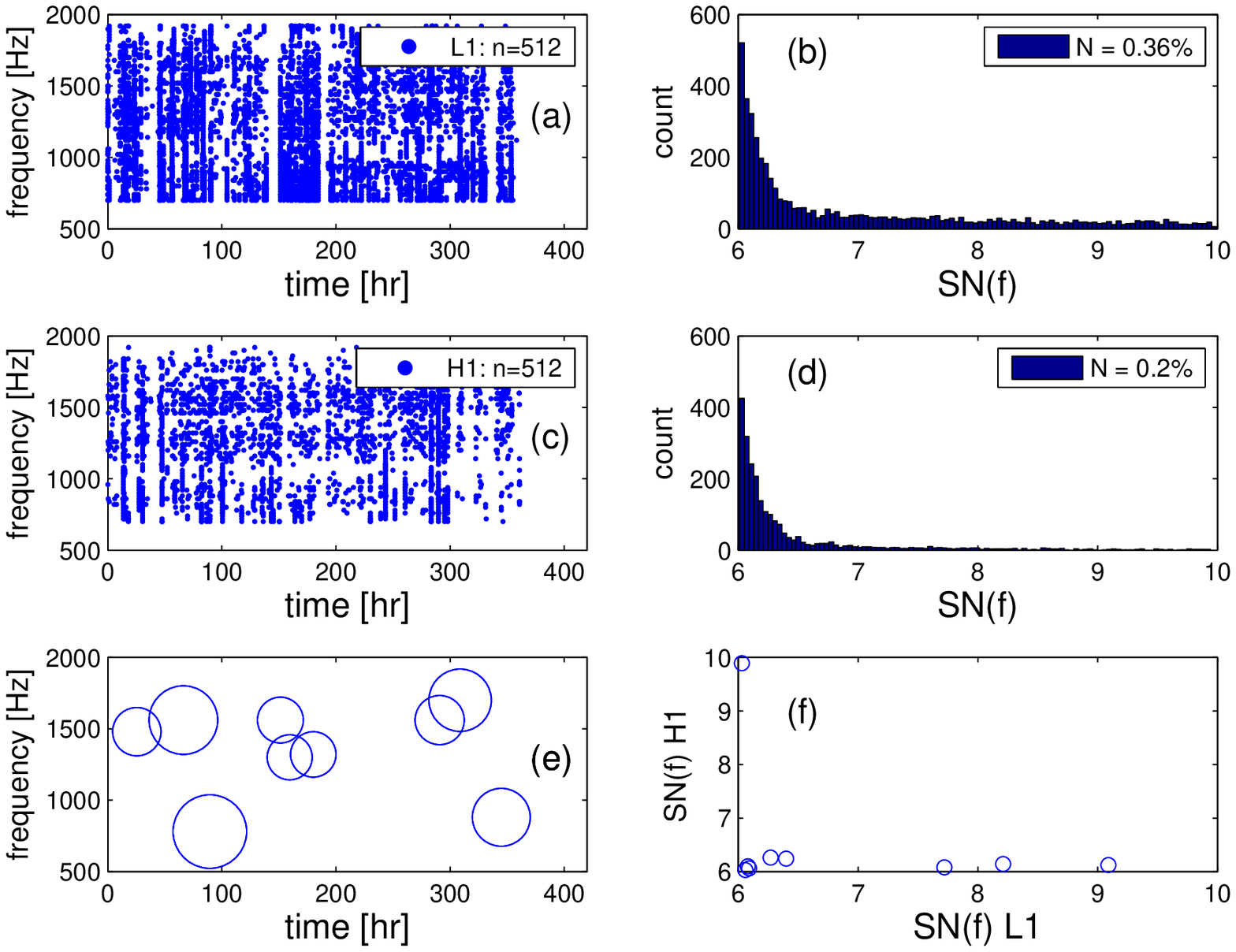}\includegraphics[scale=0.46]{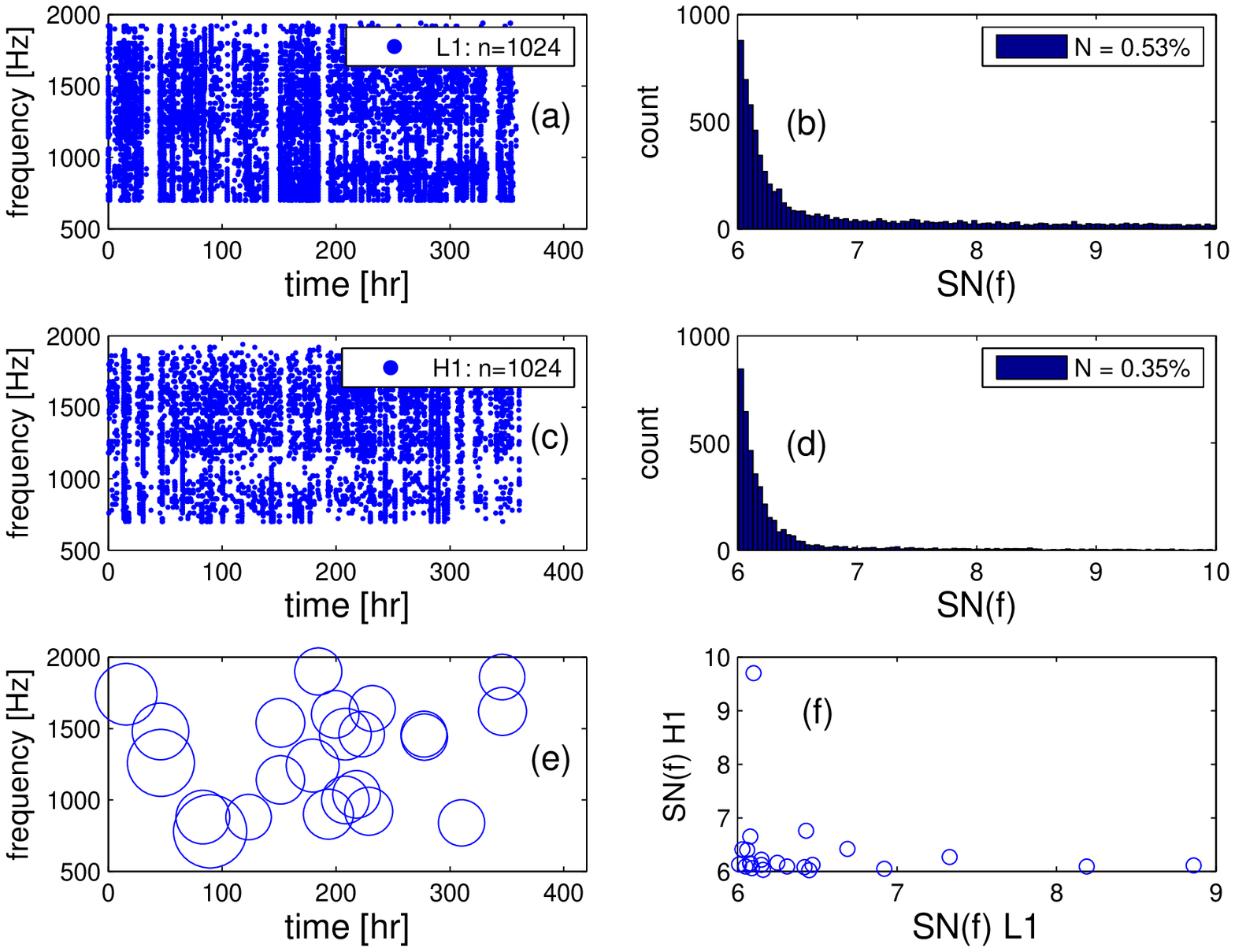}} 
\caption{Scatter plots of the tails SN($f$)$>6$ events in one minute epochs of L1 (a,b) and H1 (c,d) covering two weeks of data in LIGO S6, representing a few tenths of \% of the data. Shown are results for 512 ({\em left}) and 1024 ({\em right}) scalings in $T$ in (\ref{EQN_ft}), giving template banks of 32 and, respectively, 64 thousand chirps of $\tau=1$ s duration by time slicing. Coincidences in these tails reduced to 9 ({\em left}) and 23 ({\em right}) events (e), where circles indicate the product of the SN($f$) values in L1 and H1. However, there are no multiple coincidences in any one minute epoch, nor is there a correlation between the SN($f$) of L1 and H1 in (f). These coincidences, coarse grained at one minute resolution, appear to be insignificant.}
\label{fig6}
\end{figure}

BEGE originates from a secular evolution of the putative inner engine of a core-collapse supernova event. It hereby
describes broadband emissions, that cover a certain frequency range. A sufficiently strong signal hereby should appear
in both L1 and H1 with a correlation in the SN($f$) over a certain frequency bandwidth. BEGE can hereby be searched
for by first selecting individual events from L1 and H1 in the tail of their distributions of SN($f$), followed by a search for coincidences. Fig. \ref{fig6} shows a scatter plot of SN($f$)$>$6 in each detector, 
that comprises a few tenths of percent of the data.
Coincidences on the time scale of minutes reduces this to about 9 events out of a total of about 20,000 minutes covering two weeks of data. However, there are no multiple events covering a certain bandwidth within any one minute epoch, 
that might suggest a genuine BEGE signal. Nor is there are correlation in these 9 SN($f$) values from L1 and H1, 
even though L1 and H1 are very similar. For this reason, we these coincidences appear to be spurious events of instrumental origin (including signal injection tests).
 
\section{Conclusion and Outlook}


Modern robot optical surveys promise to provide a wealth of nearby core-collapse supernova events,
that provide attractive opportunities for directed searches of their potential emission in gravitational waves
associated with new born neutron stars and black holes. In particular, Type Ib/c supernovae stand out for 
producing black holes more likely so than neutron stars by their association to LGRBs, calorimetry 
on some of the hyper-energetic events, and the seamless unification of LGRBs and SGRB(EE)s by black hole
inner engines. Their event rate is about two orders of magnitude more numerous than the latter, making any 
gravitational wave emission from fall back matter onto rotating black holes attractive opportunities for LIGO-Virgo and KAGRA. At a distance of about 10Mpc, SN 2010br provides a challenging example of an exceptionally nearby event.

Gravitational radiation from astrophysical sources generally represents conversion of angular momentum.
Ascending and descending chirps result from conversion of orbital, respectively, spin angular momentum. 
In fall back matter onto the ISCO of a rotating black hole, they may meet and partially overlap at about
a few hundred Hz, i.e., in the shot noise dominated sensitivity bandwidth of LIGO-Virgo and KAGRA, 
and may be long lasting in representing viscous processes in accretion and feedback from the central
black hole. 

To search for BEGE, we here presented a dedicated pipeline to search in nearby events. 
As (magneto-)hydrodynamic sources of gravitational waves, phase coherence will be limited to 
intermediate time scales, representing tens or at most hundreds of wave periods associated 
with orbiting in an accretion disk. To search for long duration chirps of either sign, we apply
a butterfly filter in the time-frequency domain in the form of a bandpass for the slope $df/dt$ by
application of TSMF, using a large bank of chirp templates of intermediate duration $\tau$ each
with finite bandwidth. Here, $\tau$ is generally chosen to be on the order of 0.1 to 1 second. 

For concreteness, we here considered the nearby event SN 2010br. It 
may have been intrinsically weak or it was discovered rather late after its true time-of-onset. 
We applied our search method to two weeks of S6 data starting one month prior to 
the Type Ib/c SN 2010br. With a chirp resolution in 1024 steps, TSMF comprised 64 thousand chirp 
templates of various frequencies and bandwidths. Our two-week analysis of both L1 and H1
takes about one month computation on a 64 core mini-supercomputer by embarrassingly 
parallel computing \citep[e.g.][]{fos95}. Our task was distributed over various personal computers using a cloud 
operating system originally developed for remote sensing \citep[][]{van07}; any other software for 
distributed computing on multi-core systems \citep[e.g.][]{mig12,sin13} will serve the same purpose, 
to realize essentially optimal sensitivity by exploiting a diversity in modern supercomputing
configurations. 

Critical to a potentially successful probe is a well-determined true time-of-onset $t_0$ 
obtained from a well-sampled optical light curve, preferably with an uncertainty of less than
one week. With the poorly sampled light curve of SN 2010br, however, we face 
considerable uncertainty in $t_0$. This motivated the development of a two step
search, starting with a coarse grained TSMF to identify events of interest, here defined by SN $>$ 6.
These events are down selected by correlating the results from L1 and H1, here down to
9 and 23 events for template banks of 32 and, respectively, 64 thousand chirps. 

These particular events can be followed up by in-depth searches, i.e., (i) a two dimensional
scaling in duration and frequency of the model templates (\ref{EQN_ft}) and (ii) coincidence
analysis, by further considering time-delays between L1 and H1 detections at specific 
chirp templates that should reflect the geographic distance between the two detector sites. 
The first is exemplified in \cite{van15b}. The latter is readily included by appending to the TSMF output
the offset $\delta t$, resolved down to the sampling time of 1/4096 s in SN($t$) in the 
slices that produce the maxima SN($f$).

In the present case, however, the (economized, coarse grained) coincidence events in Fig. 8 
do not carry any signature of a common excitation of L1 and H1 by a broadband gravitational wave signal. Rather than further analysis (using a larger bank of chirp templates) of these particular events, it appears that a search epoch longer than the present two week period prior to SN2010br may be opportune. 

The method presented here is proposed as a new pipeline for systematic probes for BEGE
from nearby energetic core-collapse events provided by robotic optical surveys in the upcoming
era of advanced LIGO-Virgo and KAGRA. A detection of BEGE promises identification of their
inner engine by complete calorimetry on their energetic output.

{\bf Acknowledgments.} The author thanks the referees for constructive comments. 
This research is based in part on a Basic Research Grant (2015)
from the Korean National Research Foundation and made use of LIGO S6 data from the LIGO Open Science Center (losc.ligo.org), provided by the LIGO Laboratory and LIGO Scientific Collaboration. LIGO is funded by the U.S. National Science Foundation. Some of this work was supported by MEXT, JSPS Leading-edge Research
Infrastructure Program, JSPS Grant-in-Aid for Specially Promoted
Research 26000005, MEXT Grant-in-Aid for Scientific Research on
Innovative Areas 24103005, JSPS Core-to-Core Program, A. Advanced
Research Networks, and the joint research program of the Institute for
Cosmic Ray Research, University of Tokyo. Parallel computations were performed
using a Cloud OS VPGEONET.

{\sc \bf SUPPORTING INFORMATION}

Additional Supporting Information may be found in the online version of this article \citep{zenodo.45298}:\\
\\
{\bf ISCOWAVES}: Fortran program on bifurcation diagram ISCO waves\\
{\bf ISCOCHIRP}: MatLab program on descending ISCO chirps for injection experiments and TSMF\\

\label{lastpage}


\begin{thebibliography}{99}
\bibitem[Abbott et al.(2015)]{abb15} Abbott, B.P., Abbott, R., Abbott, T.D., et al., arXiv:1511.04398v1
\bibitem[Amati et al. (2002)]{ama02} Amati, L. et al., 2002, A\&A, 390, 81
\bibitem[Amati et al. (2006)]{ama06} Amati, L., 2006, MNRAS, 372, 233
\bibitem[Araya-G\'ochez(2004)]{ara04} Araya-G\'ochez, R.A., 2004, MNRAS, 355, 336
\bibitem[Aasi et al.(2015)]{aas15} Aasi, J., Abbott, B.P., Abbott, R., et al., 2015, ApJ, 2015, 813, 39
\bibitem[Abramovici et al.(1992)]{abr92} Abramovici, A., Althouse, W.E., Drever, R.W.P., et al., 1992, Science, 256, 325
\bibitem[Acernese et al.(2006)]{ace06} Acernese, F. et al. (Virgo Collaboration), 2006, Class. Quantum Grav., 23, S635
\bibitem[Acernese et al.(2007)]{ace07} Acernese, F. et al. (Virgo Collaboration), 2007, Class. Quantum Grav., 24, S381
\bibitem[Baiotti \& Rezolla(2008)]{bai08} Baiotti, Giacomazzo \& Rezzolla, 2008, Phys. Rev. D, 78, 084033
\bibitem[Bardeen et al.(1972)]{bar72} Bardeen, J.M., Press, W.H., \& Teukolsky, S.A., 1972, Phys. Rev. D, 178, 347 
\bibitem[Bell et al.(2015)]{bel15} AAS Meeting \#225, \#328.04
\bibitem[Bisnovatyi-Kogan(1970)]{bis70} Bisnovatyi-Kogan G. S., 1970, Astron. Zh., 47, 813
\bibitem[Bromberg et al.(2006)]{bro06} Bromberg, O., Levinson, A., \& van Putten, M.H.P.M., 2006, NewA, 619, 627
\bibitem[Burrows \& Lattimer(1987)]{bur87} Burrows, A., \& Lattimer, J.M., 1987, ApJ, 318, L63
\bibitem[Cappellaro et al.(2015)]{cap15} Cappellaro, E., Botticella, M.T., Pignata, G., et al., 2015, A\&A, 584, A62
\bibitem[Carter(1968)]{car68}Carter, B., 1968, Phys. Rev., 174, 1559
\bibitem[Cerd\'a-Dur\'an et al.(2013)]{cer13} Cerd\'a-Dur\'an, P., DeBrye, N., Aloy, M.A., et al., 2013, ApJ, 779, L18 
\bibitem[Chomiuk \& Soderberg(2010)]{cho10} Chomiuk, L., \& Soderberg, A., 2010, ATel \#2587
\bibitem[Ciufolini \& Pavlis(2004)]{ciu04} Ciufolini, I., \& Pavlis, E.C., 2004, Nature, 431, 958.
\bibitem[Ciufolini(2007)]{ciu07} Ciufolini, I., 2007, Nature 449, 41
\bibitem[Ciufolini et al.(2009)]{ciu09} Ciufolini, I., Paolozzi, A., Pavlis, E.C., et al., 2009, Space Sci Rev., 148, 71
\bibitem[Coughlin et al.(2015)]{cou15} Coughlin, M., Meyers, P., Kandhasamy, S., et al., 2015, Phys. Rev. D, 92, 43007
\bibitem[Cutler \& Thorne(2002)]{cut02} Cutler, C., \& Thorne, K. S. 2002, arXiv:gr-qc/0204090v1
\bibitem[Della Valle(2003)]{del03}Della Valle, M., Malesani, D., Benetti, S., et al., 2003, A\&A, 406, L33
\bibitem[Della Valle(2006)]{del06}Della Valle, M., 2006, Chin. J. Astron. Astrophys., 6 (Suppl), 315 
\bibitem[Della Valle(2010)]{del10}Della Valle, M., 2010, Mem. S.A. It., 81, 367
\bibitem[Diehl et al.(2006)]{die06} Diehl, R., Halloin, H., Kretch, K., et al., 2006, Nature, 439, 45
\bibitem[Drout et al.(2011)]{dro11} Drout, M.R., Soderberg, A.M., Gal-Yam, A., et al., 2011, ApJ, 741, 97
\bibitem[Duez et al.(2004)]{due04} Duez, M. D., Shapiro, S. L., \& Yo, H.-J. 2004, Phys. Rev. D, 69, 104016
\bibitem[Everitt et al.(2011)]{eve11} Everitt, C.W.F., et al., 2011, Phys. Rev. Lett. 106, 221101
\bibitem[Flanagan \& Hughes(1998)]{fla98} Flanagan, E.E., \& Hughes, S.A., 1998, Phys. Rev. D, 57, 4535
\bibitem[Foster(1995)]{fos95} Foster, I., {\em Designing and Building Parallel Programs: Concepts and Tools for Parallel Software Engineering} (Addison Wesley, Reading, MA, 1995).
\bibitem[Fox et al.(2005)]{fox05} Fox, D.B., et al., 2005, Nature, 437, 845
\bibitem[Fryer et al.(2002)]{fry02} Fryer, C. L., Holz, D.E., \&  Hughes, S. A., 2002, ApJ, 565, 430
\bibitem[Fryer \& Kimberly(2011)]{fry11} Fryer, C.L., \& Kimberly, C.B., 2011, Living Rev. Relativity, 14, 1 
\bibitem[Gammie(2001)]{gam01} Gammie, C. F., 2001, ApJ, 553, 174
\bibitem[Gehrels et al.(2005)]{geh05} Gehrels, N., Nature, 437, 851
\bibitem[Gossan et al.(2015)]{gos15} Gossan, S.E., Putton, P., Stuver, A., et al., 2015, arXiv:1511.02836v1
\bibitem[Guetta \& Della Valle(2007)]{gue07} Guetta, D., Della Valle, M.,, 2007, ApJ, 657, 73
\bibitem[Hadley \& Fernandez(2014)]{had14} Hadley, K.Z., \& Fernandez, P., 2014, Astrophys. Space Sci., 353, 191
\bibitem[Heo et al.(2015)]{heo15} Heo, J.-E., Yoon, S., Lee, D.-S., et al., 2016, NewA, 42, 24
\bibitem[Hj\"orth et al.(2005)]{hjo05} Hj\"orth, J., et al., 2005, Nature, 437, 859
\bibitem[KAGRA(2014)]{kag14} KAGRA Project (NAOJ), http://gwcenter.icrr.u-tokyo.ac.jp/en/
\bibitem[Kobayashi et al.(1997)]{kob97} Kobayashi, S., Piran, T., \& Sari, R., 1997, ApJ, 490, 92
\bibitem[Kobayashi \& Meszaros(2003)]{kob03} Kobayashi, S., \& Meszaros, P. 2003, ApJ, 589, 861
\bibitem[Kouveliotou et al.(1993)]{kov93} Kouveliotou, C., et al., 1993, ApJ, 413, L101
 \bibitem[Kulkarni(2014)]{kul14} Kulkarni, S.R., Zwicky Transient Factory Proposal, priv. commun.
\bibitem[Levinson et al.(2015)]{lev15} Levinson, A., van Putten, M.H.P.M., Pick, G., 2015, ApJ, 812, 124
\bibitem[Li et al.(2011)]{li11} Li, W., Leaman, J., Chornock, R., et al. 2011, MNRAS, 412, 1441
\bibitem[Lipunov(83)]{lip83} Lipunov, V.M., 1983, Ap\&SS, 97, 121
\bibitem[Lipunova et al.(2009)]{lip09} Lipunova, G.V., Gorbovskoy, E.S., Bogomazov, A.I., \& Liponov, V.M., 2009, MNRAS, 397, 1695
\bibitem[Lovelace \& Romanova(2014)]{lov14} Lovelace, R.V.E., \& Romanova,  M.M., 2014, Fluid Dyn. Res., 46, 041401
\bibitem[MacFayden \& Woosley(1999)]{mac99} MacFadyen A. I., \& Woosley S. E., 1999, ApJ, 524, 262
\bibitem[Mazalli et al.(2005)]{maz05} Mazzali, P.A., et al., 2005, Science, 308, 1284
\bibitem[Mejia et al.(2005)]{mej05} Mejia, A.C., et al., 2005, ApJ, 619, 1098
\bibitem[Metzger et al.(2011)]{met11} Metzger, D.B., et al., 2011, Mon. Not. R. Astron. Soc. 413, 2031
\bibitem[Mighell(2012)]{mig12} Mighell, K.J., 2012, ASP Conf. Ser. 461, eds. P. Ballester, D. Egret \& N.P.F. Lorente, p.13
\bibitem[Modjaz(2014)]{mod14} Modjaz, M., et al., 2014, AJ, 147, 99M
\bibitem[Nakar \& Piran(2002)]{nak02} Nakar, E., \& Piran, T., 2002, MNRAS, 330, 920
\bibitem[Nathanail et al.(2015)]{nat15} Nathanail, A, Strantzalis, A., \& Contopoulos, I., 2015, MNRAS, 455
\bibitem[Nevski et al.(2010)]{nev10} V. Nevski et al., 2010, CBET \# 2245
\bibitem[Ott(2009)]{ott09} Ott, C. D., 2009, Classical Quantum Gravity 26, 063001
\bibitem[Paczynski(1998)]{pac98} Paczy\'nski, B.P., 1998, ApJ, 494, L45
\bibitem[Peters \& Mathews(1963)]{pet63} Peters, P.C., \& Mathews, J., 1963, Phys. Rev., 131, 435, 4479
\bibitem[Piran \& Sari(1997)]{pir97} Piran, T., \& Sari, R., 1997, arXiv:9702093
\bibitem[Piro \& Ott(2011)]{pir11} Piro, A.L., \& Ott, C.D., 2011, ApJ, 736, 108
\bibitem[Piro \& Thrane(2012)]{pir12} Piro, A.L., \& Thrane, E., 2012, ApJ, 736, 108
\bibitem[Piro \& Pfahl(2007)]{pir07} Piro, A.L., \& Pfahl, E., 2007, ApJ, 658, 1173 
\bibitem[Prestegard \& Thrane(2012)]{pre12} Prestegard, T., \& Thrane, E., 2009, https://dcc.ligo.org/public/0093/L1200204/001/burstegard.pdf  
\bibitem[Rees et al.(1974)]{ree74} Rees, M. J., Ruffini, R., \& Wheeler, J. A. 1974, {\em Black Holes, Gravitational Waves and Cosmology: An Introduction to Current Research} (New York: Gordon \& Breach), Ch.7
\bibitem[Reichert et al.(2001)]{rei01} Reichert, D. E., Lamb, D. Q., Fenimore, E. E., Ramirez-Ruiz, E., Cline, T. L., Hurley, K., 2001, ApJ, 552, 57
\bibitem[Rice et al.(2005)]{ric05} Rice, W.K.M., Lodato, G., \& Armitage, P.J., 2005, MNRAS, 364, L56
\bibitem[Sari \& Piran(1997)]{sar97} Sari, R., \& Piran, T., 1997, ApJ, 485, 270
\bibitem[Scolnic et al.(2011)]{pan11} Scolnic, D., Riess, A., Huber, M., et al., 2011, AAS, \#218, 127.09
\bibitem[Shahmoradi \& Nemiroff(2015)]{sha15} Shahmoradi, A., \& Nemiroff, R.J., 2015, MNRAS, 451, 126
\bibitem[Singh et al.(2013)]{sin13} Singh, N., Browne, L.-M., \& Butler, R., 2013, Astron. Comput., 2, 1
\bibitem[Skakura \& Sunyaev(1973)]{sha73} Shakura, N.I., \& Sunyaev, R.A., 1973, Astron. Astophys., 24, 337
\bibitem[Smartt(2009)]{sma09} Smartt, S. J., 2009, ARA\&A, 47, 63
\bibitem[Soderberg et al.(2008)]{sod08}Soderberg, A.M., Berger, E., Page, K.L., et al. 2008, Nature, 453, 469
\bibitem[Soderberg et al.(2010)]{sod10}Soderberg, A.M., Chakraborti, S., Pignata, G., et al. 2010, Nature, 463, 513
\bibitem[Somiya et al.(2012)]{som12} Somiya, K., (for the KAGRA Collaboration), 2012, Class. Quantum Grav., 29, 124007 
\bibitem[Sutton et al.(2010)]{sut10} Sutton, P.J., Jones, G., Chatterji, S., et al., 2010, N. J. Phys.,  12,053034
\bibitem[Tagger et al.(1990)]{tag90} Tagger, M., Henriksen, R.N., Sygnet, J.F., \& Pellat, R., 1990, ApJ 353, 654
\bibitem[Tagger \& Pellat(1999)]{tag99} Tagger, M., \& Pellat, R., 1999, A\&A, 349, 1003
\bibitem[Tagger(2001)]{tag01} Tagger, M., 2001, A\&A, 380, 750
\bibitem[Tagger \& Varni(2006)]{tag06} Tagger, M., \& Varni\`ere, P., 2006, ApJ, 642, 1457
\bibitem[Tagger \& Melia(2006)]{tag06b} Tagger, M., \& Melia, F., 2006, ApJ, 636, L33
\bibitem[Taubenberger(2009)]{tau09} Taubenberger, S., et al., 2009, MNRAS, 397, 677
\bibitem[Taylor \& Weisberg(1989)]{tay89} Taylor, J.H., \& Weisberg, J.M., 1989, ApJ, 345, 434
\bibitem[Taylor(1994)]{tay94} Taylor, J.H., 1994, Rev. Mod. Phys., 66, 711
\bibitem[Thompson(1994)]{tho94} Thompson, C., 1994, Mon. Not. R. Astron. Soc. 270, 480
\bibitem[Thorne(1980)]{tho80} Thorne, K.S., 1980, Rev. Mod. Phys., 52, 299
\bibitem[Thrane et al.(2011)]{thr11} Thrane, E., Kandhasamy, S., Ott, C.D., et al., 2011, Phys. Rev. D, 83, 083004
\bibitem[Thrane \& Coughlin(2013)]{thr13} Thrane, E., \& Coughlin, M., 2013, Phys. Rev. D 88, 083010
\bibitem[Thrane \& Coughlin(2014)]{thr14} Thrane, E., \& Coughlin, M., 2014, Phys. Rev. D 89, 063012
\bibitem[Usov(1994)]{uso94} Usov, V., 1994, MNRAS, 267, 1035
\bibitem[Weisberg et al.(2010)]{wei10} Weisberg, J.M., Nice, D.J., \& Taylor, J.H., 2010, ApJ, 722, 1030
\bibitem[Vallisneri et al.(2014)]{val14} Vallisneri, M., Kanner, J., Williams, R., Weinstein, A., \& Stephens, B., 2014, $in$ Proc. 10th LISA Symposium, University of Florida, Gainesville, May 18-23; http://dx.doi.org/10.7935/K5RN35SD
\bibitem[van Putten \& Eardley(1996)]{van96} van Putten, M.H.P.M., \& Eardley, D.M., 1996, Phys. Rev. D, 53, 3056
\bibitem[van Putten(1999)]{van99} van Putten, M.H.P.M., 1999, Science, 284, 115
\bibitem[van Putten(2000)]{van00} van Putten, M.H.P.M., 2000, Phys. Rev. Lett., 84, 3752
\bibitem[van Putten(2001)]{van01a} van Putten, M.H.P.M., 2001, Phys. Rev. Lett., 87, 091101
\bibitem[van Putten \& Ostriker(2001)]{van01} van Putten, M.H.P.M., \& Ostriker, E., 2001, ApJ, 552, L31
\bibitem[van Putten(2002)]{van02} van Putten, M.H.P.M., 2002, ApJ, 575, L71
\bibitem[van Putten \& Levinson(2003)]{van03} van Putten, M.H.P.M., \& Levinson, A., 2003, ApJ, 584, 937
\bibitem[van Putten(2004)]{van04} van Putten, M.H.P.M., 2004, ApJ, 611, L81
\bibitem[van Putten et al.(2004)]{van04b} van Putten, M.H.P.M., Levinson, A., Lee, H.K., Regimbau, T., Punturo, M., \& Harry, G.M., 2004, Phys. Rev. D, 2004, 69, 044007 
\bibitem[van Putten, \& van Putten(2007)]{van07} van Putten, M.H.P.M., \& van Putten, A.F.P., 2007, Proc. R. Soc. A, 463, 2495
\bibitem[van Putten(2008a)]{van08} van Putten, M.H.P.M., 2008a, ApJ, 684, L91
\bibitem[van Putten(2008b)]{van08b} van Putten, M.H.P.M., 2008b, ApJ, 685, L63
\bibitem[van Putten \& Gupta(2009)]{van09} van Putten, M.H.P.M., \& Gupta, A.C., 2009, MNRAS, 394, 2238
\bibitem[van Putten et al.(2011)]{van11} van Putten, M.H.P.M., Kanda, N., Tagoshi, H., Tatsumi, D., Masa-Katsu, F., \& Della Valle, M., 2011, Phys. Rev. D 83, 044046
\bibitem[van Putten et al.(2011a)]{van11a} van Putten, M.H.P.M., Della Valle, M., \& Levinson, A., 2011a, A\&A, 535, L6
\bibitem[van Putten(2012)]{van12} van Putten, M.H.P.M., 2012, Prog. Theor. Phys., 127, 331
\bibitem[van Putten \& Levinson(2012)]{van12b} van Putten, M.H.P.M., \& Levinson, A., 2012, {\em Relativistic Astrophysics of the Transient Universe} ) (Cambridge University Press)
\bibitem[van Putten(2013)]{van13} van Putten, M.H.P.M., 2013, Acta Polytechnica, 53 (Suppl.), 736
\bibitem[van Putten et al.(2014)]{van14} van Putten, M.H.P.M., Guidorzi, C., \& Frontera, P., 2014, ApJ, 786, 146
\bibitem[van Putten et al.(2014)]{van14b} van Putten, M.H.P.M., Gyeong-Min, Lee, Della Valle, M., Amati, L., \& Levinson, A., 2014, MNRASL, 444, L58
\bibitem[van Putten(2014)]{van14c} van Putten, M.H.P.M., 2014, MNRAS, 447, L11  
\bibitem[van Putten(2015a)]{van15a}{van15a}  van Putten, M.H.P.M., 2015, MNRAS, 447, L11
\bibitem[van Putten(2015b)]{van15b} van Putten, M.H.P.M., 2015, ApJ, 810, 7
\bibitem[van Putten(2016)]{zenodo.45298} van Putten, M.H.P.M., 2016, {Broadband chirps from ISCO waves around rotating black holes}, Zenodo, doi:10.5281/zenodo.45298   
\bibitem[Villasenor et al.(2005)]{vil05} Villasenor, J.S., et al., 2005, Nature, 437, 855
\bibitem[Wen \& Schutz(2005)]{wen05} Wen, L., \& Schutz, B.F., 2005, Class. \& Quant. Gravity 22, 1321
\bibitem[Woosley(1993)]{woo93} Woosley, S.L., 1993, ApJ, 405, 273
\bibitem[Woosley(2006)]{woo06} Woosley, S.E., \& Bloom, J.S., 2006, ARA\&A., 44, 507
\end{thebibliography}
\end{document}